\documentclass[aps,pra,twocolumn] {revtex4-1}
\usepackage{amsmath,amssymb,graphicx,amsfonts}

\usepackage{times}
\usepackage[colorlinks=true,linkcolor=blue,urlcolor=blue,citecolor=blue,breaklinks=true]{hyperref}

\usepackage[utf8]{inputenc}
\DeclareUnicodeCharacter{03BC}{\mbox{$\mu$}}
\DeclareUnicodeCharacter{2009}{ }

\def\wt{\widetilde} 

\usepackage{float}

\def\eq#1{Eq.~(\ref{#1})}

\def\da{\dagger}

\def\da{\dagger}

\date\today
\begin{document}
	
\title{Local spectral density of an interacting one-dimensional Bose gas  with an impurity}

\author{Aleksandra Petkovi\'{c}}

\affiliation{Laboratoire de Physique Th\'{e}orique, Universit\'{e} de Toulouse, CNRS, UPS, 31062 Toulouse, France}

\begin{abstract}
A weakly interacting Bose gas with a single static impurity is studied in one dimension under the assumption that its phase coherence length is longer than the system size. The local spectral density is evaluated and an analytic expression valid for positive strengths of the impurity coupling to the bosons, at all distances and at all frequencies is obtained using the Bogoliubov-de Gennes approximation. \end{abstract}

\maketitle

\section{Introduction}

The influence of an impurity particle on a quantum many-body bath leads to many interesting phenomena in physics, such as the Kondo effect \cite{Kondo}, the orthogonality catastrophe  \cite{orthogonalityCatastrophe}, and the insulating behavior of one-dimensional (1D) electron systems \cite{KaneAndFisher}. 
A high degree of control in the manipulation and fabrication of ultracold atomic gases has renewed the interest in impurity physics by allowing us to simulate problems of condensed matter physics in correlated fermionic \cite{PhysRevLett.102.230402,PhysRevLett.103.170402,NaturePolarons,NaturePolaronsII,PhysRevLett.118.083602,Cetina96,NatureCasimir} and bosonic environments \cite{chikkatur2000suppression,palzer2009quantum,zipkes2010trapped,schmid2010dynamics,PhysRevLett.109.235301,catani2012quantum,weitenberg2011single-spin,fukuhara2013quantum,PhysRevLett.117.055302,Meinert945}. 
In this work we consider a 1D system containing a single impurity coupled to an interacting gas of bosons. Some properties of such a 1D system have been studied previously, such as the friction force exerted on the impurity \cite{castro_neto1996dynamics,astrakharchik2004motion,gangardt2009bloch,PRLimpurity,PhysRevB.101.104503},  the effective impurity mass \cite{fuchs2005spin,PhysRevA.104.052218}, the dynamic correlation functions \cite{zvonarev2007spin,matveev2008spectral,lamacraft2009dispersion,kamenev2009dynamics}, the impurity relaxation dynamics \cite{burovski2014momentum,knap2014quantum,10.21468/SciPostPhys.8.4.053}, and the boson density profile \cite{PhysRevResearch.2.043104}. 

Here, we focus on the local spectral density \cite{PhysRevB.74.085114} of bosons, that has not been considered yet. It is defined as 
 \begin{align}\label{nwdef}
 n(x,\omega)=-\frac{1}{\pi} \text{Im}[\mathcal{G}(x,x;\omega)] ,
 \end{align}
with the time-ordered Green's function
 \begin{align}\label{G}
 \mathcal{G}(x,x';\omega)=-i\int_{-\infty}^{\infty} dt e^{i(\omega+\mu/\hbar) t}\langle \mathcal{T}[\hat\Psi(x,t)\hat\Psi^\dag(x',0)]\rangle.
 \end{align} 
Here, $\mathcal{T}$ denotes the time-ordering, $\hat\Psi(x,t)$ is the bosonic annihilation field operator, and $\hbar$ is the reduced Planck constant. The ensemble average is denoted by $\langle\ldots\rangle$.  Note that the energy $\hbar \omega$ is measured with respect to the chemical potential $\mu$ of the bosons. 
The local spectral density carries important information about the system. It is proportional to the transition rate for adding and removing a boson with energy $\hbar\omega$ at position $x$, where the proportionality constant depends on the frequency as $1+e^{-(\hbar\omega+\mu)/T}$ and $1+e^{(\hbar\omega+\mu)/T} $, respectively \cite{PhysRevB.74.085114}. We set the Boltzmann constant to unity and $T$ is the temperature. At zero temperature, $n(x,\omega)/\hbar$ coincides with the local single-particle density of states. 

Many experimental techniques, which allow us to characterize many-body ultracold gases, were developed recently. For example, the single-particle spectral function and the dynamic structure factor of 1D bosonic atoms can be measured by the inelastic light scattering, i.e., the Bragg spectroscopy \cite{PhysRevLett.83.2876,PhysRevLett.82.4569,PhysRevA.72.023407,Clement,PhysRevA.79.043623,BraggSpectroscopy}, while their local correlation functions can be probed by measuring photoassociation rates \cite{PhysRevLett.95.190406} and particle losses \cite{PhysRevLett.92.190401,PhysRevLett.107.230404}. 
The most direct way to measure the local spectral density is via the cold atom tunneling microscope \cite{PhysRevA.76.063602}. This probe is analogous to the scanning tunneling microscope for condensed matter systems.
Powerful imaging techniques employing the quantum gas microscope \cite{Nature2009} or the ion-based microscope \cite{Stecker_2017,PhysRevX.11.011036} also could be used. 
\begin{figure}
\includegraphics[width=\columnwidth]{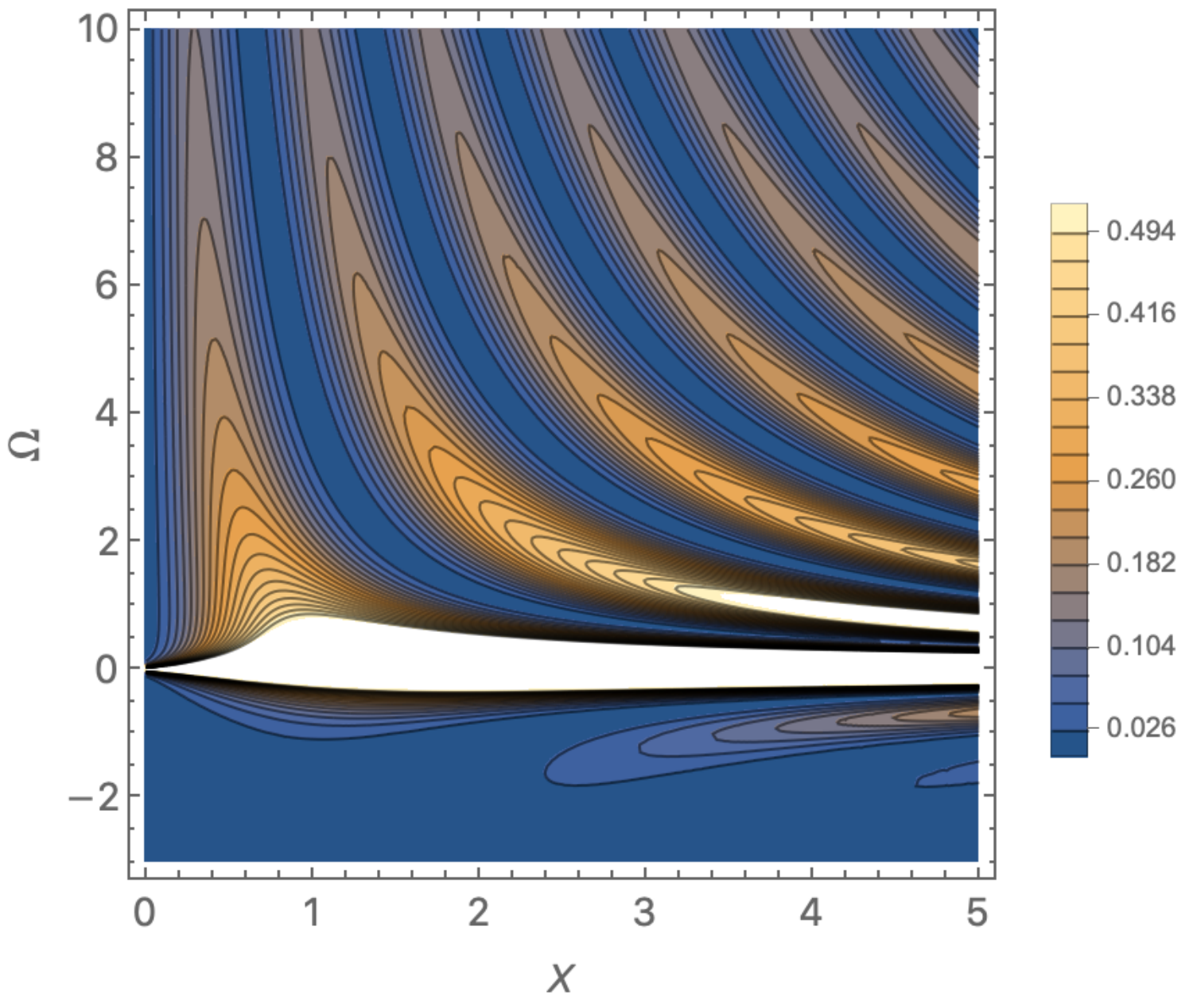}
\caption{(Color online) Dimensionless quantity $I(X,\Omega)$ as a function of scaled position $X=x/\xi$ and frequency $\Omega=\hbar\omega/\mu$  determines the local spectral density of the bosons, see \eq{nwn}. The impurity with strong dimensionless coupling constant $\widetilde{G}=10$ is located at $X=0$. The temperature is set to $T=\mu/10$. }
\label{Fig:contour}
\end{figure}

The influence of an impurity on correlation functions of interacting 1D fermions has attracted a great deal of attention \cite{KaneAndFisher,PhysRevLett.71.3351, PhysRevB.51.17827,PhysRevLett.76.1505,PhysRevLett.75.3505}. At zero temperature, a single impurity cuts a system of repulsively interacting fermions into two disconnected parts  leading to the reduction of the local spectral density at its position \cite{KaneAndFisher}. As a result, the power-law exponent of the frequency dependence of the local spectral density in the vicinity of the impurity differs from that in the bulk at low energies \cite{KaneAndFisher,PhysRevB.51.17827,PhysRevLett.76.1505}. 

In this work we use a microscopic description of weakly interacting bosons within the Bogoliubov-de Gennes approximation. We evaluate the local spectral density and obtain an analytic expression valid for positive strength of the impurity coupling to the bosons and at all distances and frequencies. The scattering of Bogoliubov quasiparticles at the impurity leads to oscillations of the local spectral density, see Fig.~\ref{Fig:contour}. The oscillations persist at an arbitrary small impurity-boson coupling constant, as well as far from the impurity. Moving away from the impurity,  the function $n(x,\omega)$ becomes periodic in $x$ for a given frequency $\omega$ and $|x|\gg \xi$, where $\xi$ is the healing length. Its period is $\pi/k$, where $\hbar k$ is the momentum of the Bogoliubov quasiparticle with energy $\hbar \omega$. The average value of $ n(x,\omega)$ over the period equals the local spectral density in the absence of the impurity. 

At zero temperature, we find that $n(x,\omega)\propto1/ |\omega|$ at low frequency, for any impurity coupling constant.
Note that this singular behavior holds also in the absence of the impurity. This conclusion is in agreement with a renormalization-group prediction that an impurity with a finite coupling constant is an irrelevant perturbation in a system of bosons with local interactions \cite{KaneAndFisher}. We point out that even in the case of an impurity with an infinitely strong coupling constant that cuts the system into two disconnected parts (i.e., the boundary of the system), the power-law exponent of the frequency dependence of $n(x,\omega)$ in the vicinity of the impurity remains unchanged from the bulk one.  This result holds in the limit of weakly interacting bosons. Increasing the boson-boson interaction strength, the exponent of the frequency dependence of $n(x,\omega)$ in the bulk becomes different from the one in the vicinity of the boundary in the limit of low energies \cite{Cazalilla2}.

The paper is organized as follows. After introducing the model in Sec.~\ref{model}, we solve the equation of motion for the single particle operator in Sec.~\ref{sec:solution}. A general expression for the local spectral density is given in Sec.~\ref{sec:LSD}. The system without the impurity is considered in Sec.~\ref{sec:homogene}.  In the limit of low frequencies, the local spectral density for an arbitrary impurity strength is considered  in Sec.~\ref{sec:small}, while Sec.~\ref{LargeOmega} is devoted to high frequencies.  The case of an infinitely strong impurity coupling constant is studied in Sec.~\ref{boundary} at arbitrary frequencies.  The opposite limit of a weakly coupled impurity is presented in Sec.~\ref{sec:weakly}. 
The case of arbitrary impurity strength and frequency is studied in Sec.~\ref{sec:arbitrary}. We conclude the paper in Sec.~\ref{sec:conclusions}. Some results are relegated to appendices.

%%%%%%%%%%%
\section{Model\label{model}}
%%%%%%%%%%%%

We study the effects of a static impurity on a 1D system of interacting bosons. The system is modeled by the Hamiltonian
\begin{align}
H=\int dx\left(-\hat\Psi^\da \dfrac{\hbar^2\partial_x^2}{2m}\hat\Psi+\dfrac{g}{2}\hat\Psi^\da\hat\Psi^\da\hat\Psi\hat\Psi\right)+G \hat\Psi^\da(0)\hat\Psi(0).\label{H}
\end{align}
Here $g>0$ denotes  the repulsive contact interaction between bosons with mass $m$. We assume that the mean density of the bosons is $n_0$. We introduce the dimensionless parameter $\gamma=gm/\hbar^2n_0$ that measures the strength of the interaction. The bosonic field operators satisfy the commutation relations $[\hat\Psi(x),\hat\Psi^\da(x')]=\delta(x-x')$ and $[\hat\Psi(x),\hat\Psi(x')]=0$, with $\delta(x)$ being the Dirac delta function. The impurity is situated at position $x=0$. The strength of the coupling between the impurity and the Bose gas is denoted by $G$.  We study weakly interacting bosons, $\gamma\ll 1$, while the impurity--boson coupling is of arbitrary strength. 
The static impurity can be experimentally realized with heavy impurity atoms or with obstacles. The latter could be created, e.g., by optical defects in the form of a tightly focused laser beam transversely oriented relative to the 1D condensate. 

In order to analyze the local spectral density, we first solve the equation of motion for the single-particle field operator. It is given by $i\hbar \partial_t\hat\Psi=[\hat\Psi,H]$. After introducing the dimensionless coordinate $X=x\sqrt{m\mu}/\hbar$ and  time $\tau=t\mu/\hbar$, the field operator can be rewritten as $\hat\Psi(x,t)=\sqrt{\mu/g}\hat\psi(X,\tau)e^{-i\tau}$. The chemical potential is denoted by $\mu$. The field operator $\hat\psi(X,\tau)$ can be expanded as \cite{pitaevskii_bose-einstein_2003,sykes,Casimir}
\begin{align}\label{expansion}
\hat\psi(X,\tau)=\psi_0(X)+\alpha \hat\psi_1(X,\tau)+\alpha^2{\psi}_2(X)+\mathcal{O}(\alpha^3),
\end{align}
where the small parameter is $\alpha=(\gamma g n_0/\mu)^{1/4}\approx \gamma^{1/4}\ll 1$. The time independent function $\psi_0$ describes the condensate wave function in the absence of fluctuations. The time-dependent field operator $\hat\psi_{1}$ accounts for the quantum and thermal fluctuations. The time-independent function $\psi_2$ accounts for corrections of the condensate wave function $\psi_0$ caused by fluctuations \cite{sykes,Casimir}. The chemical potential $\mu$ can be expressed as a function of the mean boson density $n_0$ using $n_0=\int dx \langle \hat\Psi^{\dag}(x,t) \hat\Psi(x,t) \rangle /L$ and Eq.~(\ref{expansion}). Here, $L$ denotes the system length. We obtain an expansion of the chemical potential in $\alpha$. In the leading order,  $\mu=g n_0$ \cite{lieb_exact_1963}, assuming $L\gg \xi$. Here, $\xi=1/n_0\sqrt{\gamma}$ is the healing length.

Note that for one spatial dimension long-wavelength phase fluctuations destroy the condensate in the thermodynamic limit even at zero temperature. At zero temperature, the phase coherence length is $\ell_{\phi}=\xi \exp{(2\sqrt{\pi^2/\gamma})}\gg \xi$ \cite{petrov_low-dimensional_2004}.  At sufficiently high temperatures, the phase coherence length is determined by the temperature as $\ell_{\phi}=\hbar^2 n_0/m T$ \cite{PhysRevLett.91.040403,PhysRevA.67.053615}. The perturbative expansion of the field operator (\ref{expansion}) holds for $L \ll \ell_{\phi}$. Thus, our approach is applicable under the condition $\ell_{\phi}\gg L\gg \xi$, which implies that the temperature satisfies $T\ll \hbar^2 n_0^2 \sqrt{\gamma} /m\approx\mu/\sqrt{\gamma}$.

%%%%%%%%%%%%%%%%%%%%%%%%%%%
\section{Solution of the equation of motion\label{sec:solution}}
%%%%%%%%%%%%%%%%%%%%%%%%%%%

Substituting the expansion (\ref{expansion}) into the equation of motion we obtain one equation for each power of $\alpha$. To the zeroth order in $\alpha$ we get
\begin{align}
{\widehat{\mathcal{L}}}_1(X)\psi_0(X)=0,\label{eom0}
\end{align}
where the operator  ${\widehat{\mathcal{L}}}_j(X)$ is
\begin{align}\label{operator}
{\widehat{\mathcal{L}}}_j(X)=-\frac{\partial^2_X}{2}+j |\psi_0(X)|^2-1+\widetilde{G}\delta(X).
\end{align}
We introduced the dimensionless parameter $\wt G$ that measures the strength of the impurity coupling to the bosons as
\begin{align}\label{tildeG}
\wt G=G\sqrt{m/\mu}/\hbar.
\end{align}
We are interested in a solution of \eq{eom0} with a vanishing superfluid current far from the impurity.  We assume that the boson density remains unperturbed far from the impurity. We are not interested in the boundary effects, but rather focus on the impurity influence. Thus, we do not impose vanishing density at the boundaries of the system. The finite system length $L$ introduces an infrared cutoff in the theory, and prevents the long-wavelength fluctuations from destroying the condensate. However, some results hold in the thermodynamic limit as well, as we discuss below. 

In the absence of the impurity, \eq{eom0} is known as the Gross-Pitaevskii equation, and its solution is $\psi_0^{\textrm{hom}}(x)=1$.
The solution of \eq{eom0} is given by \cite{KovrzinMaximov}
\begin{align}\label{Psi0}
\psi_0(X)=\tanh(|X|+X_0),
\end{align}
where parameter $X_0$ carries the information about the impurity strength and is given by
\begin{align}\label{xo}
X_0=\tanh^{-1}\left[\dfrac{1}{2}\left(-\wt G+\sqrt{4+\wt G^2}\right)\right].
\end{align}
This expression follows from the condition imposed on the derivative of the wave function at the position of the impurity, $\psi_0'(0^+)-\psi_0'(0^-)=2\widetilde{G} \psi_0(0)$.
For vanishing coupling constant $\wt{G}$, the parameter $X_0$ tends to infinity and we obtain $\psi_0^{\textrm{hom}}(X)$. For large coupling constant, $\wt{G}\to \infty$, the parameter $X_0\to 0$, and the wave function becomes $\psi_0(X)\to \tanh{|X|}$ leading to a vanishing boson density at the impurity position.

The effects of thermal and quantum fluctuations to first order in $\alpha$ are described by $\hat\psi_1$. Its equation of motion reads
\begin{align}
i\partial_\tau\hat\psi_1(X,\tau)=&{\widehat{\mathcal{L}}}_2(X)\hat\psi_1(X,\tau)+\psi_0^2(X)\hat\psi_1^\da(X,\tau). \label{eom1}
\end{align}
We search for the solution of \eq{eom1} in the form \cite{pitaevskii_bose-einstein_2003}
\begin{align}\label{ps1}
\hat\psi_1(X,\tau)=\sum_k N_k\left[u_k(X)\hat{b}_ke^{-i\epsilon_k \tau}-v^*_k(X)\hat{b}_k^\da e^{i\epsilon_k \tau}\right], 
\end{align}
where $N_k$ is a normalization constant to be defined below. 
The bosonic operators describing the Bogoliubov quasiparticles $\hat{b}_k$ and $\hat{b}_k^\da$ satisfy the bosonic commutation relations $[\hat{b}_k,\hat{b}^\da_{k'}]=\delta_{k,k'}$ and $[\hat{b}_k,\hat{b}_{k'}]=0$. 
Substituting \eq{ps1} in \eq{eom1} we obtain two coupled equations for $u_k$ and $v_k$ known as the Bogoliubov-de Gennes equations
\begin{align}
\epsilon_k u_k(X)={}&{\widehat{\mathcal{L}}}_2(X)u_k(X)-\psi_0^2(X)v_k(X), \label{eq:bdguv1}\\
-\epsilon_kv_k(X)={}&{\widehat{\mathcal{L}}}_2(X) v_k(X)-\psi_0^2(X)u_k(X). \label{eq:bdguv2}
\end{align}
We will first solve these equations without the $\delta$-potential in the operator ${\widehat{\mathcal{L}}}_2(X)$. Then, we will take it into account through the conditions $u_k(0^+)=u_k(0^-)$ and $u'_k(0^+)-u'_k(0^-)=2\widetilde{G} u_k(0)$. The same constraints are to be imposed on $v_k(X)$. The prime denotes the derivative with respect to $X$.

We decouple Eqs.~(\ref{eq:bdguv1}) and (\ref{eq:bdguv2}) by introducing the functions $S(k,X)=u_k(X)+v_k(X)$ and $D(k,X)=u_k(X)-v_k(X)$, which satisfy fourth-order differential equations. The equation for $S(k,X)$ has the form
\begin{align}\label{eqS}
\epsilon_k^2S(k,X)={\widehat{{L}}}_3(X)\widehat{{{L}}}_1(X) S(k,X),
\end{align}
while $D(k,X)$ can be expressed in terms of $S(k,X)$ as
\begin{align}\label{Dk}
D(k,X)=\dfrac{1}{\epsilon_k}\widehat{{L}}_1(X) S(k,X).
\end{align}
Here, we introduced $\widehat{L}_j=-{\partial^2_X}/{2}+j |\psi_0(X)|^2-1$. 
The four independent solutions of Eqs.~(\ref{eqS}) and (\ref{Dk}) are given by \cite{Kovrzin}
\begin{align}
S_n(k,X)=&\left\{ik_n-2\tanh \left[X+\text{sgn}(X)X_0\right]\right\}e^{i k_n X},\label{S}\\
D_n(k,X)=&\dfrac{ik_n}{\epsilon_{k_n}}\frac{1}{\cosh^{2}\left(\left|X\right|+ X_0\right)}e^{i k_n X}+\dfrac{k_n^2}{2\epsilon_{k_n}}S_n(k,X),\label{D}
\end{align}
where $n\in \{1,2,3,4\}$ and the energy dispersion reads as
\begin{align}
\epsilon_k=\sqrt{k^2+k^4/4}.
\end{align}
The four roots of the energy dispersion entering in $S_n$ are $k_{1,2}=\pm k$ and $k_{3,4}=\pm i\sqrt{4+k^2}$ in terms of 
$
k=\sqrt{2}\sqrt{\sqrt{\epsilon_k^2+1}-1}
$.

Now we are ready to account for the existence of the $\delta$-potentials in Eqs.~(\ref{eq:bdguv1}) and (\ref{eq:bdguv2}). The general solution for $S(k,X)$ is a linear combination of the four solutions, which for $k>0$ reads
\begin{align}\label{solution}
&S(k,X)=
\begin{cases}
 S_1(k,X)+ r S_2(k,X)+r_e S_4(k,X),& X<0\\
t S_1(k,X)+t_e S_3(k,X),& X>0,
\end{cases}
\end{align}
while $D(k,X)$ has the form
\begin{align}
&D(k,X)=
\begin{cases}
 D_1(k,X)+ r D_2(k,X)+r_e D_4(k,X),& X<0\\
t D_1(k,X)+t_e D_3(k,X),& X>0.
\end{cases}
\end{align}
Equation (\ref{solution}) describes an incoming wave from $X=-\infty$ with $k>0$ that is partially reflected from the impurity at the origin. In  \eq{solution} we omitted from the linear combination the unphysical exponentially growing solutions, i.e., $S_3$ for $X<0$ and $S_4$ for $X>0$.
The four remaining coefficients $r,r_e,t$, and $t_e$ in \eq{solution} are to be determined such that the solutions for $S(k,X)$ and $D(k,X)$ are continuous and their derivative with respect to $X$ are discontinuous functions at the impurity position:  
\begin{gather}
S(k,0^+)=S(k,0^-),\label{e1}\\
D(k,0^+)=D(k,0^-),\\
\partial_X S(k,0^+)-\partial_X S(k,0^-)=2\wt G S(k,0),\\
\partial_X D(k,0^+)-\partial_X D(k,0^-)=2\wt G D(k,0).\label{e4}
\end{gather} 
We evaluate the coefficients, finding
\begin{widetext}
\begin{align}
&r=\frac{i k \left(1-\eta ^2\right) \left[k^2 (2 \eta +q)+4 \left(\eta ^3+\eta \right)+q^3+2 \eta  q^2+4 \eta ^2 q\right]}{\left[k \eta +i \left(\eta ^2+1\right)\right] (k-i q) \left[k^2 (2 \eta +q)+i k (2 \eta +q)^2-2 \eta  \left(2 \eta ^2+q^2+2 \eta  q-2\right)\right]},\label{r}\\
&t=\eta\frac{2 k^2 \left(k^2+4\right) \eta +q^3 \left(k^2+2 \eta ^2+2\right)+2 \eta  q^2 \left(k^2+2 \eta ^2+2\right)+q \left[k^4+2 k^2 \left(\eta ^2+1\right)+4 \eta ^4-4\right]}{\left[k \eta +i \left(\eta ^2+1\right)\right] (k-i q) \left[k^2 (2 \eta +q)+i k (2 \eta +q)^2-2 \eta  \left(2 \eta ^2+q^2+2 \eta  q-2\right)\right]},\label{t}\\
&t_e=r_e={}\frac{4 k \eta  \left(\eta ^2-1\right)}{(k-i q) \left[k^2 (2 \eta +q)+i k (2 \eta +q)^2-2 \eta  \left(2 \eta ^2+q^2+2 \eta  q-2\right)\right]}\label{Re}.
\end{align}
\end{widetext}
Here, we introduced $\eta=\left(-\wt G+\sqrt{4+\wt G^2}\right)/2$ and $q=\sqrt{4+k^2}$. The reflection and the transmission probability for quasiparticles are given by $|r|^2$ and $|t|^2$, respectively. They are non monotonic functions of momenta.
 
For negative $k$-values we should consider the scattering process of an incoming wave from $X=\infty$. However, since the impurity is static, we can use the property $S(-k,X)=S(k,-X)$ and use the solution (\ref{solution}) for $k>0$. 

The normalization constant $N_k$  in \eq{ps1} follows from the requirement \cite{pitaevskii_bose-einstein_2003}
\begin{align}\label{normalisation}
N_k N_q\int_{-L/2\xi_{\mu}}^{L/2\xi_{\mu}}d X[u_k(X)u_q^*(X)-v_k(X)v_q^*(X)]=\delta_{k,q}.
\end{align}
where 
\begin{align}
\xi_{\mu}=\hbar/\sqrt{m\mu}.
\end{align}
%Note that the dimensionless position $X$ is actually measured in units of $\xi_{\mu}$, i.e., $X=x/ \xi_{\mu}$. 
We evaluate $N_k=(\xi_\mu/2L\epsilon_k)^{1/2}$ in the limit $L\gg \xi_{\mu}$. We point out that $N_k$ does not depend on the impurity strength  $\widetilde{G}$. The solution of the equation of motion calculated here will be analyzed and simplified in some limiting cases in the following sections.

%%%%%%%%%%%%%%%%%%%%%%%%
\section{Local spectral density}\label{sec:LSD}
 %%%%%%%%%%%%%%%%%%%%%%%%

%%%%%%%%%%%%%%%%%%%
\subsection{Physical meaning of the local spectral density}
%%%%%%%%%%%%%%%%%%%

At zero temperature the local spectral density (\ref{nwdef}) can be represented as
\begin{align}\label{nT0}
n(x,\omega)=&\sum_{\zeta_{N+1}}|\langle \zeta_{N+1}|\hat\Psi^\dag(x,0)|0\rangle |^2\delta(\omega-\epsilon_{\zeta_{N+1}}/\hbar)\notag\\&+\sum_{\zeta_{N-1}}|\langle \zeta_{N-1}|\hat\Psi(x,0)|0\rangle |^2 \delta(\omega+\epsilon_{\zeta_{N-1}}/\hbar),
\end{align}
where $|0\rangle$ denotes the ground state of the system of $N$ bosons, and  the excitation energy $\epsilon_{\zeta_{N}}=E_{\zeta_{N}}-E_0(N)\geq 0$. Here $E_{\zeta_{N}}$ is the energy of the $N$-particle state $|\zeta_{N}\rangle$ and $E_0(N)$ denotes the ground state energy of the system with $N$ bosons. Note that states $|\zeta_N\rangle$ form a complete set of eigenstates of the Hamiltonian (\ref{H}) for $N$ particles. The chemical potential  at zero temperature is $\mu=E_0(N+1)-E_0(N)$. In Eq.~(\ref{nT0}), we have assumed $N\gg 1$. We thus see that $n(x,\omega)/\hbar$ at zero temperature denotes the local single-particle density of states. 

\begin{widetext}
At nonzero temperature, the local spectral density (\ref{nwdef}) takes the form
\begin{align}\label{nN}
n(x,\omega)=&{}\sum_{\zeta_{N+1},\zeta_N} \frac{e^{-\beta E_{\zeta_N}}}{Z}|\langle \zeta_{N+1}|\hat\Psi^\dag(x,0)|\zeta_N\rangle |^2 \delta(\omega+\frac{\mu+E_{\zeta_N}-E_{\zeta_{N+1}}}{\hbar})\notag\\ &+\sum_{\zeta_{N-1},\zeta_N}\frac{e^{-\beta E_{\zeta_N}}}{Z}|\langle \zeta_{N-1}|\hat\Psi(x,0)|\zeta_{N}\rangle |^2 \delta(\omega+\frac{\mu+E_{\zeta_{N-1}}-E_{\zeta_N}}{\hbar}) ,
\end{align}
\end{widetext}
where $Z$ is the canonical partition function of the system of $N$ bosons and the inverse temperature is denoted by $\beta=1/T$.
For $N\gg 1 $, Eq.~(\ref{nN}) can be rewritten as 
\begin{align}\label{nw1}
n(x,\omega)=&{}\frac{1}{Z}(1+e^{-\beta(\hbar\omega+\mu)})\sum_{\zeta_{N+1},\zeta_{N}}|\langle \zeta_{N+1}|\hat\Psi^\dag(x,0)|\zeta_{N}\rangle |^2\notag\\ & \times \delta[\omega+(\mu+E_{\zeta_{N}}-E_{\zeta_{N+1}})/\hbar] e^{-\beta E_{\zeta_N}}.
\end{align}
From Eqs.~(\ref{nN},\ref{nw1}) it follows that the local spectral density is proportional to the transition rate for adding and removing a boson with energy $\hbar \omega$ at position $x$, where the proportionality constant depends on the frequency as $1+e^{-\beta(\hbar\omega+\mu)}$ and $1+e^{\beta(\hbar\omega+\mu)} $, respectively. The energy $\hbar \omega$ is measured with respect to $\mu$. 

%%%%%%%%%%%%%%%%%%%%
\subsection{Evaluation of the local spectral density}
%%%%%%%%%%%%%%%%%%%%

Here, we evaluate the local spectral density by employing the expansion (\ref{expansion}) in Eq.~(\ref{nwdef}). We can write $n(x,\omega)$ in the form
\begin{align}\label{nW}
n(x,\omega)=&2\frac{\mu}{g}\delta(\omega)\left\{\psi^2_0(x/ \xi_{\mu})+2\alpha^2 \psi_0(x/ \xi_{\mu})\text{Re} [\psi_2(x/ \xi_{\mu})]\right\}\notag\\&+
\alpha^2\frac{\hbar}{g}I(x/ \xi_{\mu},\hbar\omega /\mu)+\mathcal{O}(\alpha^3).
\end{align}
Here we took into account that $\psi_0$ is a real function. We evaluate the contribution to $n(x,\omega)$ originating from $\hat\psi_1$  to be the dimensionless
\begin{align}
\label{Iw}
I(X,\Omega)=&\frac{1}{4\pi}\frac{1}{k_{\Omega}\sqrt{1+\Omega^2}} 
 \coth{\left(\frac{|\Omega|\mu}{2 T}\right)}
\Big[\theta{(\Omega)}\big(|u_{k_{\Omega}}(X)|^2\notag\\&+|u_{-k_{\Omega}}(X)|^2\big)+\theta{(-\Omega)}\big(|v_{k_{\Omega}}(X)|^2\notag\\&+|v_{-k_{\Omega}}(X)|^2\big)\Big],
\end{align}
where the dimensionless $k_{\Omega}$ is defined as 
\begin{align}\label{kOmega}
 k_{\Omega}=\sqrt{2}\sqrt{\sqrt{\Omega^2+1}-1},
\end{align}
and $\theta{(x)}$ denotes the Heaviside theta function. We distinguish two contributions in \eq{Iw}, one from the hole-like excitations in the case of negative frequencies and another for the particle-like excitations defined for positive frequencies. We used the expressions  $\int_{0}^{\infty}e^{i s t} dt=\lim_{\delta\to 0^+}i/(s+i \delta)$ and $\lim_{\delta\to 0^+}\frac{1}{\pi}\frac{\delta}{x^2+\delta^2}=\delta(x)$. 

Quantum and thermal fluctuations bring particles out of the condensate and are taken into account  through the field operator $\hat{\psi}_1$ and the function ${\psi}_2$.  The latter describes the depletion of the condensed particles caused by the fluctuations \cite{sykes,Casimir} and thus brings no new frequency dependence in $n(x,\omega)$.
Notice that thermal effects enter \eq{Iw} through the average occupation numbers of Bogoliubov quasiparticles $n_p=\langle \hat{b}^{\dag}_p \hat{b}_p\rangle=1/[\exp{( \epsilon_p\mu/T)}-1]$ leading to the multiplicative factor $\coth({|\Omega|\mu/2 T})=1+2 n_{k_{\Omega}}$. 
%The quantity $n_p$ has to be distinguished with respect to the occupation number of the bare bosonic particles. 

Expressing the chemical potential $\mu$ in terms of the mean boson density $n_0$ as $\mu=g n_0$ in the leading order \cite{lieb_exact_1963}, we rewrite the local spectral density as
\begin{align}\label{nwn}
n(x,\omega)= 2 n_0\psi_0^2(x/\xi)\delta{(\omega)}+\frac{1}{v}I\left(\frac{x}{\xi},\frac{\hbar\omega}{\mu}\right).
\end{align}
Here $v$ is the sound velocity and has the form $v=\sqrt{gn_0/m}$. The function $I(X,\Omega)$ is given by \eq{Iw}, $\xi=1/n_0\sqrt{\gamma}$ and $\psi_0$ is given by \eq{Psi0}.
The condensed particles have zero momentum and thus determine the contribution proportional to $\delta(\omega)$ in Eq.~(\ref{nwn}). The factor multiplying $\delta(\omega)$ is written in the leading order, i.e., the contribution originating from $\psi_2$ has been omitted. The reason is that in the present work we focus on the $\omega$ dependence of the local spectral density and $\psi_2$ does not bring any new $\omega$ dependence. Furthermore, Eq.~(\ref{nwn}) holds in the parameter region where the true condensate exists (see the discussion in Sec.~\ref{model}). There the contribution from $\psi_2$ in the local spectral density is negligible. 
	
For a quasi-condensate the system length is longer than the phase coherence length and the temperature is assumed to be low, $T\ll \mu/\sqrt{\gamma}$, such that the density fluctuations remain small with respect to the mean-field value of the local density \cite{petrov_low-dimensional_2004,PhysRevA.67.053615}. The result (\ref{nwn}) certainly holds for a quasicondensate for high enough frequencies $\omega\gg v/\ell_{\phi}$ where the decoherence effects are not visible. Nevertheless, at low frequencies $\omega\ll \mu/\hbar\gg  v/\ell_{\phi}$ the Luttinger liquid description \cite{Giamarchi,Haldane} indicates that the finite frequency part of \eq{nwn} remains valid for arbitrarily small frequency (see Appendix \ref{AppA}).

Replacing the functions $u_k$ and $v_k$ in Eq.~(\ref{Iw}) by the expressions presented in the preceding section, we obtain the local spectral density for any impurity coupling constant $\widetilde{G}>0$. For the case of attractive impurity-boson interaction, we restrict our analysis to moderate values of $|\tilde{G}|$ due to the collapse of the bosons onto the impurity at strong attraction. In what follows we evaluate the local spectral density in some limiting cases where it acquires a simple analytic form. In the other parameter regions we analyze it numerically. 

%%%%%%%%%%%%%%%%%%%%%%%%%%%%%%%%%%%%%
\section{Homogeneous system $\widetilde{G}=0$}\label{sec:homogene}
%%%%%%%%%%%%%%%%%%%%%%%%%%%%%%%%%%%%%

In the absence of the impurity ($\widetilde{G}=0$) the Hamiltonian (\ref{H}) is known as the Lieb-Liniger model \cite{lieb_exact_1963,lieb1963exactII}. Many correlation functions characterising this exactly solvable model have been studied. The long distance and time asymptotic behavior of dynamic correlation functions was studied using the form factor approach \cite{Kitanine_2012}, while  the operator product expansion is used to derive high energy and momentum asymptotics \cite{sekino2020field}. The nonlinear Luttinger liquid  theory provided the exponents that characterise the singularities of  dynamic correlation functions in the vicinity of the edges \cite{RevModPhys.84.1253}, while the algebraic Bethe ansatz allowed the study of response functions numerically for large finite-size systems \cite{PhysRevA.74.031605,Caux_2007}. 
In this section, as a special case of our problem, we study a uniform weakly interacting Bose gas within the standard Bogoliubov-de Gennes  approximation \cite{pitaevskii_bose-einstein_2003} that allows us to evaluate the local spectral density at arbitrary frequencies.

In the absence of the impurity, the expression (\ref{Iw}) becomes
\begin{align}\label{additional}
I^{\textrm{hom}}(\Omega)=\frac{ \sqrt{\Omega ^2+1}+ \Omega}{2\sqrt{2} \pi \sqrt{\left(\Omega
   ^2+1\right) \left(\sqrt{\Omega ^2+1}-1\right)}} \coth{\left(\frac{|\Omega|\mu}{2T}\right)}.
\end{align}
Here, we have introduced the  superscript $^{\textrm{hom}}$ to point out that the system is homogeneous.
We remind the reader that in \eq{additional} one must use $\mu=g n_0$. Replacing $\psi_0$ by $\psi_0^{\textrm{hom}}=1$ and function $I(X,\Omega)$ by \eq{additional} in Eq.~(\ref{nwn}) we obtain the local spectral density in a homogeneous system. The functions $u_k(x)$ and $v_k(x)$ are given in Appendix \ref{App:homogene}.

%%%%%%%%%%%%%%%%%%%%%
%\subsection{Zero temperature}
%%%%%%%%%%%%%%%%%%%%%%

We first consider the zero temperature case.
Note that thermal effects in \eq{additional} enter through the multiplicative factor that becomes unity at zero temperature. The function $I^{\textrm{hom}}$ is $X$-independent, as expected for a translationally invariant system. Its asymmetry for positive and negative $\Omega$ originates from the difference of particle-like and hole-like contributions in \eq{Iw} determined by the expressions for $u_k^{\textrm{hom}}$ and $v_k^{\textrm{hom}}$ given by Eqs.~(\ref{homogeneU}) and (\ref{homogeneV}), respectively.
Nevertheless, their first order term in the expansion in small $k$ is the same.  The local spectral density is divergent at the chemical potential since $I^{\textrm{hom}}(\Omega)$ in the limit $|\Omega|\ll 1$ reads as 
\begin{align}\label{wToO}
I^{\textrm{hom}}(\Omega)=\frac{1}{2\pi |\Omega |}.
\end{align}
The reason is that the interaction leads to the depletion of the condensate bringing the bosons mostly into the low-energy states. 
In the condensate of the length $L$, the infrared cutoff $\Omega_{\mathrm{min}}=\hbar \omega_{\mathrm{min}}/\mu$ is of the order of $\xi_{\mu}/L\ll 1$. Thus, the number of particles out of the condensate remains finite and negligible \footnote{Note that the local density is given by $n(x)=\lim_{t\to0^-}i \int_{-\infty}^{\infty} e^{-i\omega t}G(x,x;\omega)d\omega/2\pi$.
}.

At high energies the behavior of particle and hole contributions is very different. For high positive energy $\Omega\gg 1$ it is given by 
\begin{align}\label{BigPositive}
I^{\textrm{hom}}(\Omega)=\frac{1}{\pi \sqrt{2 \Omega}},
\end{align}
while for large negative energy $-\Omega\gg 1 $, it behaves as 
\begin{align}\label{BigNegative}
I^{\textrm{hom}}(\Omega)=\frac{1}{4\sqrt{2}\pi}\frac{1}{|\Omega|^{5/2}}.
\end{align}
%The function $I^{\textrm{hom}}(\Omega)$ is the increasing function for negative and the decreasing function for positive $\Omega$.

%%%%%%%%%%%%%%%%%%%%%%%
%\subsection{Finite temperature}
%%%%%%%%%%%%%%%%%%%%%%%

Next we consider the effects of thermal fluctuations on the spectral density. From  \eq{additional} it follows that only for energies well above the temperature $\hbar |\omega|\gg T$ (or equivalently $|\Omega|\gg T/\mu$) thermal fluctuations become exponentially suppressed and can be neglected, thus leading to the zero temperature result. Otherwise, thermal fluctuations do affect the spectral density. Namely, in the limit of low energies, $|\Omega|\ll \min\{T/\mu,1\}$, the divergent behavior is observed, 
\begin{align}\label{IhomOmegasmall}
I^{\textrm{hom}}(\Omega)=\frac{T}{\pi \Omega^2\mu }+\mathcal{O}(\Omega^{-1}),
\end{align}
which is to be contrasted with \eq{wToO}. In the case of $\mu/\sqrt{\gamma}\gg T \gg \mu$ and for  intermediate energies $T/\mu\gg\Omega\gg 1$, we get 
\begin{align}
I^{\textrm{hom}}(\Omega)=\frac{1}{\pi}\sqrt{\frac{2}{ \Omega^3}}\frac{T}{\mu },
\end{align}
while  for $T/\mu\gg-\Omega\gg 1$ we get 
\begin{align}\label{LargeNOmegaT}
I^{\textrm{hom}}(\Omega)=\frac{1}{2\sqrt{2}\pi}\frac{T}{|\Omega|^{7/2}\mu}.
\end{align}

In Appendix \ref{AppA} we use the Luttinger liquid theory to evaluate  the local spectral density for an arbitrary interaction strength  between the bosons, at finite temperatures, and  in the thermodynamic limit. The final result is given by \eq{nwanyg}. Note that this approach provides only the low-energy description of our system.  For weakly interacting bosons, it thus gives only the low-frequency $\hbar|\omega|\ll \mu$ behavior of the local spectral density given by  $I^{\textrm{hom}}(\hbar\omega/\mu)/v= \sqrt{\gamma}{n_0}\coth{(\hbar\omega /2T)}/{2\pi \omega}$. Thus, the singular behavior (\ref{wToO}) holds in the thermodynamic limit as well.

%%%%%%%%%%%%%%%%%%%%%%
\section{Low frequency and arbitrary impurity strength\label{sec:small}}
%%%%%%%%%%%%%%%%%%%%%%

\subsection{Low frequency $\hbar|\omega|\ll \mu$}

Here, we consider the low frequency behavior of the local spectral density, $\hbar|\omega|\ll \mu$. The case of very strong impurity coupling constant $\hbar|\widetilde{G}\omega|/\mu \gtrsim1$ needs to be considered separately, as we explain in the following section. 

The energy dispersion of Bogoliubov quasiparticles takes a phonon-like form at low energies, and the impurity barrier becomes almost transparent, i.e., the reflection amplitude (\ref{r}) is $r=0$ to leading order \cite{KovrzinMaximov}. Note that this result is very different from the problem of a free single particle scattering from a $\delta$-function potential. In the latter case the reflection probability becomes one in the limit of low energies \cite{Griffiths}. 

In Appendix \ref{LowFrequency}, we give simplified expressions for the functions $u_k(X)$ and $v_k(X)$ at small $|k|\ll 1$ .
Now we can evaluate $I(X,\Omega)$ given by \eq{Iw}. It takes the form
\begin{align}\label{IsmallOmega}
I(X,\Omega)=
%&\frac{1}{8 \pi  (\eta +1)^2 \Omega } \text{sech}^4(\left|
%   X\right| +X_0) \coth \left(\frac{\mu  \Omega }{2
  % T}\right) \notag\\
   %&\times\Big|(\eta -1) e^{2 X_0}+(\eta +1)e^{-i\Omega
     %\left|  X\right|}\{1 -i \notag\\&\times  \sinh [2
   %(\left| X\right| +X_0)]\}\Big|^2\\
   &\frac{1}{8 \pi \Omega }  \coth \left(\frac{\mu  \Omega }{2
   T}\right) \mathrm{sech}^4(\left|
   X\right| +X_0)\notag\\
   &\times\Big|1 -i  \sinh [2
   (\left| X\right| +X_0)] -e^{i\Omega
     \left|  X\right|} \Big|^2
\end{align}
to leading order in small frequency.
We remind the reader that $X_0$ is given by \eq{xo}. We can now use Eq.~(\ref{nwn}) with $I(X,\Omega)$ given by \eq{IsmallOmega} to evaluate the local spectral density. 

From \eq{IsmallOmega} follows that the singular behavior observed in the absence of the impurity $\sim \coth \left({\mu  \Omega }/{2T}\right)/\Omega$ remains unchanged in the presence of the impurity. As expected, the impurity introduces an additional $X$-dependence of the local spectral density. It causes the oscillations that are characterized by the new characteristic length $\xi \mu/\hbar\omega$ in the limit of low frequencies. We discuss these oscillations in more detail for arbitrary frequencies in the following sections. 
However, note that the oscillations at low frequencies (i.e., in \eq{IsmallOmega}) are not visible and  we can approximate 
\begin{align}\label{SmallOmega1}
I(X,\Omega)\approx \frac{1}{2\pi \Omega}{\coth \left(\frac{\mu  \Omega }{2T}\right)}\tanh^2{(|X|+X_0)}.
\end{align}
Increasing the impurity coupling strength $\widetilde{G}$, the local spectral density decreases.
%In the limit of weak impurity $|\widetilde{G}|\ll 1$ the parameter $X_0\approx \log{(4/G)}/2$ and we obtain $I= \coth \left({\mu  \Omega }/{2T}\right)\{1+e^{-2|X|}\widetilde{G}[\sin(\Omega |X|)-1]\}/2\pi \Omega$ in agreement with \eq{additional}. 
For any $\widetilde{G}$, the local spectral density at large separations $x\gg \xi$ and low frequency has the same dependence as  in the absence of the impurity, i.e., considering $|X|\gg 1$ in \eq{IsmallOmega} we recover the homogeneous case. 
%We notice that the result (\ref{IsmallOmega}) is not symmetric function of $\Omega$ in a general case, regardless the fact it describes the low-energy limit.

%%%%%%%%%%%%%%%%%%%%%%%%%%%%%%%%%%%%%%
\subsection{Large coupling $\widetilde{G}\gg 1$ and low frequency $\hbar|\omega|\ll \mu$}\label{OmegaSGL}
%%%%%%%%%%%%%%%%%%%%%%%%%%%%%%%%%%%%%%

In this subsection we consider the limit of low frequency and large positive $\widetilde{G}$ such that $\widetilde{G}|\Omega|$ can be arbitrarily small or big. Here $\Omega=\hbar\omega/\mu$. In order to evaluate $I(X,\Omega)$ in this region of parameters,  we first calculate the coefficients $t=1/(ik\widetilde{G}-1)$, $r=\widetilde{G}k/(i+\widetilde{G}k)$, and $t_e=-k/2(i+\widetilde{G}k)$ for $k=|\Omega|\ll 1$. The reflection probability $|r|^2$ vanishes as $(k\widetilde{G})^2$ for $k\widetilde{G}\ll1$ and tends to one in the opposite case $k\widetilde{G}\gg1$. Thus in the latter case of very strong impurity coupling constant, the barrier almost completely reflects the quasiparticles. 

We can now further simplify the expressions for the functions $u_k(X)$ and $v_k(X)$. They are presented in Appendix \ref{app:BigGSmallOmega}. Here we evaluate the function $I(X,\Omega)$ given by \eq{Iw} for arbitrary $X$ to be
\begin{align}\label{IGk}
I(X,\Omega)=&\frac{\mathrm{sech}^4(X) \coth \left(\frac{\mu  \Omega }{2T}\right) }{16 \pi\Omega  \left(\widetilde{G}^2\Omega ^2+1\right)}\Big\{ \Big|  i \sinh(2 |X|)+1\notag\\&-e^{-i |X| \Omega }\Big| ^2+\Big| e^{i |X|\Omega } \widetilde{G} \Omega  [\sinh(2 |X|)-i]\notag\\&+e^{-i |X| \Omega }(\widetilde{G} \Omega +i)  [\sinh (2|X|)+i]+1\Big| ^2\Big\}.
\end{align}
We point out that in the presence of an impurity with strong coupling constant, the oscillations are pronounced and remain at long separations from it. For $X\gg 1$, $I(X,\Omega)$ becomes a periodic function in $X$ and its average value (over the period) coincides with the one in the absence of the impurity $I^{\mathrm{hom}}(\Omega)$. We show in Sec.~\ref{sec:arbitrary} that the latter property holds for arbitrary frequency and arbitrary coupling constant $\widetilde{G}$.

Considering \eq{IGk} in the limit of small $\widetilde{G}|\Omega|\ll 1$, we recover the expansion of \eq{IsmallOmega} for large $\widetilde{G}\gg 1$.  Analyzing the expression (\ref{IGk}) in the opposite limit of large $\widetilde{G}|\Omega|\gg 1$ we obtain
\begin{align}\label{IbT0}
I(X,\Omega)=& \frac{1}{4 \pi   \Omega} \coth \left(\frac{\mu   \Omega  }{2 T}\right)\notag\\ &\times{\left[2 \tanh (X) \cos (X \Omega )+\text{sech}^2(X) \sin (X
   \Omega )\right]^2}.
\end{align}
in the leading order. Equation (\ref{IbT0}) describes an impurity with infinitely strong coupling constant  that cuts the system into two disconnected parts leading to the complete reflexion of the quasiparticles.  The local spectral density vanishes at the impurity position $X=0$. Nevertheless, note that the impurity does not change the power-law exponent of the frequency dependence of $n(x,\omega)$ of the homogeneous case (\eq{wToO} and \eq{IhomOmegasmall}), but rather introduces a multiplicative factor that carries $X$-dependence. This factor is an oscillating function of $\Omega$. In  the vicinity of the impurity, i.e., $X\ll1$ the behavior is $I(X,\Omega)=\coth \left(\frac{\mu   \Omega  }{2 T}\right)X^2/\pi\Omega$.
In Sec.~\ref{boundary} we consider an infinitely strong
 impurity coupling constant in more detail and evaluate the local spectral density for arbitrary frequencies. 

At low energies, one can apply  the Luttinger-liquid model for bosons in the presence of a boundary \cite{Cazalilla2}. This approach does not allow one to determine  the prefactors, but it shows that by increasing the strength of the interaction between the bosons, the difference between the power-law exponent of the frequency in $n(x,\omega)$ in the bulk  $-1+1/(2K)$ \cite{Giamarchi} and in the vicinity of an impurity with an infinitely strong repulsive coupling $-1+1/K$ \cite{Cazalilla2}, becomes visible. Here $K$ denotes the Luttinger liquid parameter, while the temperature is set to zero. For weakly interacting bosons  $K=\pi/\sqrt{\gamma}\gg 1$.

%%%%%%%%%%%%%%%%%%%%%%%%%%%%%%%%
\section{High frequency and arbitrary impurity strength \label{LargeOmega}}
%%%%%%%%%%%%%%%%%%%%%%%%%%%%%%%%

%%%%%%%%%%%%%%%%%%%%%%%%%%%%%%%%
\subsection{High frequency $\hbar|\omega|\gg\mu$}
%%%%%%%%%%%%%%%%%%%%%%%%%%%%%%%%

In this section we evaluate the local spectral density in the limit of high frequencies $|\Omega|\gg 1$, where $\Omega=\hbar\omega/\mu$. For $k\gg 1$ the transmission coefficient of Bogoliubov quasiparticles across the impurity potential becomes $|t|^2\approx1$ and  $r_e=0$. This is expected, since we discuss the particle regime of Bogoliubov quasiparticles where $\epsilon(k)=k^2/2$ \cite{Griffiths}. We provide simplified expressions for functions $u_k$ and $v_k$ in Appendix \ref{HighFrequency}. For high positive frequency $\Omega\gg 1$,  \eq{Iw} coincides with the homogeneous case
\begin{align}\label{largeOmegaP}
I(X,\Omega)=\frac{\coth \left(\frac{\mu  \Omega }{2 T}\right)}{ \pi 
   \sqrt{2\Omega }}
\end{align}
to leading order,
while for large negative frequency $-\Omega\gg 1$ it takes the form
\begin{align}\label{largeOmegaN}
I(X,\Omega)=\frac{ \coth \left(\frac{\mu\left| \Omega \right| }{2 T}\right)}{4 \sqrt{2} \pi  \left|\Omega \right| ^{5/2}} \tanh ^4(\left| X\right| +X_0).
\end{align}
For negative frequencies, the leading order result depends on the impurity strength in contrast to the case of positive frequencies. However, the influence of the impurity vanishes exponentially quickly at separations longer than the healing length. The presence of the impurity does not change the exponents of the power-law frequency dependence $n(x,\omega)$ observed in the homogeneous case \eq{additional}. In the case of strong impurity coupling constant $|\widetilde{G}|\gtrsim \sqrt{|\Omega|}\gg 1$, the above result does not apply and special attention should be paid, as we discuss in the following subsection.

%%%%%%%%%%%%%%%%%%%%%%%%%%%%%%%
\subsection{High frequency $\hbar|\omega|\gg \mu$ and large coupling $\widetilde{G}\gg 1$}\label{OmegaGL}
%%%%%%%%%%%%%%%%%%%%%%%%%%%%%%%

In this subsection we consider the limit of high frequency $|\Omega|\gg 1$ and strong impurity coupling constant $\widetilde{G}\gg 1$. For $k\gg 1$, the coefficients simplify to $t=k/(k+i \widetilde{G})$, $r=\widetilde{G}/(\tilde G-i k)$, and $r_e = 0$ in agreement with the scattering of noninteracting particles from a $\delta$-function potential \cite{Griffiths}. We provide simplified expressions for $u_k(x)$ and $v_k(x)$  in Appendix \ref{HighFrequencyB}. Finally,  \eq{Iw} becomes 
\begin{align}\label{allomega}
I(X,\Omega)=&\frac{
   \coth \left(\frac{\mu  | \Omega | }{2 T}\right)}{
   \pi  \sqrt{2| \Omega |}  \left(
   \widetilde{G}^2+2 | \Omega |\right)}\Big[2 \widetilde{G}^2 \sin ^2\left( X \sqrt{2|\Omega| }\right)\notag\\&+ \widetilde{G}
   \sqrt{2|\Omega| } \sin \left(2
   \sqrt{2|\Omega| } |X|\right)+2 |\Omega| \Big]\notag\\&\times 
 \left(\theta(\Omega)+\theta(-\Omega) \frac{\tanh ^4(X)}{4\Omega^ 2}\right).
\end{align}
Here $\theta(x)$ denotes the Heaviside step function. For large $|X|\gg1$, one should replace $\sqrt{2|\Omega|}$ by $k_{\Omega}$ ( it is given by \eq{kOmega}) inside each sine in \eq{allomega}, as well in \eq{largeGOmegaP} and \eq{largeGOmegaN}.
Notice that an impurity with strong coupling constant $\widetilde{G}\gtrsim \sqrt{|\Omega|}\gg 1$  causes the oscillations in the leading order expression for $n(x,\omega)$. 

Expanding \eq{allomega} in large $\widetilde{G}\gg \sqrt{|\Omega|}$ 
for positive energies, the asymptotic behavior is given by
\begin{align}\label{largeGOmegaP}
I(X,\Omega)=\frac{1}{\pi }\sqrt{\frac{2}{\Omega }}\coth \left(\frac{\mu   \Omega  }{2 T}\right)  \sin ^2\left(
   \sqrt{2\Omega }X\right),
\end{align}
while for negative energies we get
\begin{align}\label{largeGOmegaN}
I(X,\Omega)= \frac{\coth \left(\frac{\mu  \left| \Omega \right| }{2 T}\right)}{2 \sqrt{2} \pi  \left| \Omega \right|
   ^{5/2}} \tanh ^4(X) \sin ^2\left( X \sqrt{2\left| \Omega
   \right| }\right).
\end{align}
We see that the power-law dependence in $\Omega$ describing the system in the absence of the impurity, gets multiplied by a function of $X$ and $\Omega$. Note that it is an oscillating function of $\Omega$ for a given $X$, similarly to the case of low energies considered in the previous section.
Equations (\ref{largeGOmegaP}) and (\ref{largeGOmegaN}) describe an impurity with infinitely strong  coupling constant, leading to vanishing local boson density at its position.

The expansion of \eq{allomega} in the leading order in high frequencies $\sqrt{|\Omega|}\gg \widetilde{G}$ coincides with Eqs.~(\ref{largeOmegaP}) and  the leading order expansion of \eq{largeOmegaN} in $\widetilde{G}\gg1$.
The expression (\ref{allomega}) provides the behavior in different crossover regions.

%%%%%%%%%%%%%%%%%%%%%%%%%
\section{Infinitely strong impurity coupling constant \label{boundary}}
%%%%%%%%%%%%%%%%%%%%%%%%%

In this section we consider the case of an infinitely strong positive impurity coupling constant. The impurity cuts the system into two semi-infinite subsystems. It imposes the nullification of the single-particle bosonic operator at its position,
$
\hat\Psi(0)=0,
$
and thus the local boson density vanishes at $x=0$.  
Using the results of  Sec.~\ref{sec:solution} we evaluate the condensate wave function, $u_k(X)$, and $v_k(X)$ in Appendix \ref{AppInfinitely}.
We evaluate $I(X,\Omega)$ given by \eq{Iw}  to be
\begin{widetext}
\begin{align}\label{Ib}
 I(X,\Omega)= &\coth \left(\frac{\mu  \left| \Omega \right| }{2 T}\right) \frac{
   \left[\left(k_{\Omega }^2+2 \Omega \right) \tanh (X) \cos
   \left(X k_{\Omega }\right)+k_{\Omega } \sin \left(X k_{\Omega
   }\right) \left(\frac{k_{\Omega }^2}{2}+\text{sech}^2(X)+\Omega
   \right)\right]^2}{4 \pi  \Omega ^2 \sqrt{\Omega ^2+1}
   k_{\Omega }},
\end{align}
\end{widetext}
while the local spectral density takes the form
\begin{align}\label{Nwbb}
n(x,\omega)=&2 n_0\delta{(\omega)}\tanh^2{(x/\xi)}+\frac{1}{v} I\left(\frac{x}{\xi},\frac{\omega\hbar}{gn_0}\right).
\end{align}
The limiting cases of low and high frequencies of \eq{Ib} coincide with Eq.~(\ref{IbT0}) and Eqs.~(\ref{largeGOmegaP}) and (\ref{largeGOmegaN}), respectively.

We see that the local spectral density is an oscillating function of the coordinate for a given frequency, contrary to the local boson density that does not show Friedel oscillations \cite{PhysRevResearch.2.043104}. The amplitude of the oscillations depends on position $X$ and frequency $\Omega$. It vanishes at the impurity position $X=0$, as it has to be the case. Far away from the impurity $|X|\gg 1$,  the oscillations remain and $I(X,\Omega)$ becomes a periodic function of $X$ for a given frequency $\Omega$. Its spatial periodicity is given by $\pi/k_{\Omega}$ (see \eq{kOmega}) and at positions $X_n=\arctan{(-2/k_{\Omega})}/k_{\Omega}+n \pi/k_{\Omega}$ the function $I(X_n,\Omega)$ vanishes. Here $n$ is an integer such that $|X_n|\gg 1$.  After averaging $I(X,\Omega)$ over its period, we get the same value as in the absence of the impurity.

%%%%%%%%%%%%%%%%%%%%%%%%%%%%
\section{Weakly coupled impurity $|\widetilde{G}|\ll 1$\label{sec:weakly}}
%%%%%%%%%%%%%%%%%%%%%%%%%%%%

Here we consider the effects of a weakly coupled impurity with $|\widetilde{G}|\ll 1$. $\widetilde{G}$ is given by \eq{tildeG} and it can be rewritten to lowest order in $\gamma$ as 
\begin{align}\label{Ggamma}
\widetilde{G}=G\sqrt{\gamma}/g.
\end{align} 
We provide the solutions for the functions $u_k$ and $v_k$  in Appendix \ref{app:weakly}. They are obtained by simplifying the results from Sec.~\ref{sec:solution}. Now we are able to evaluate \eq{Iw}.  It takes the form 
\begin{align}\label{IG}
I(X,\Omega)=&I^{\mathrm{hom}}(\Omega)+\widetilde{G} \frac{\coth \left({\mu  |\Omega| }/{2 T}\right)}{4 \pi 
   \Omega  \left(\Omega ^2+1\right) k_{\Omega }}\notag\\
   &\times \Big[
    \Omega k_{\Omega } \left(\sqrt{\Omega
   ^2+1}+\Omega \right) \sin \left(2 \left| X\right|  k_{\Omega
   }\right)\notag\\&+2 k_{\Omega }e^{-\frac{2   \left| \Omega X\right| }{k_{\Omega }}}
   \sin \left(\left| X\right|  k_{\Omega }\right)\notag\\&-
   \sqrt{\Omega ^2+1} e^{-2 \left| X\right| }
   \left(k_{\Omega }^2+2\Omega \right)\Big],
   \end{align}
  where $I^{\mathrm{hom}}(\Omega)$  is the result (\ref{additional}).
Equation (\ref{IG}) applies for all $\Omega$ and all $X$. 
We can now evaluate Eqs.~(\ref{nwn}) and (\ref{additional}). The function $\psi_0$ is given by the expression (\ref{Psi0}). For the coherence of the presentation, we can expand it in small $\widetilde{G}$ as $\psi_0(X)=1-\frac{1}{2} \widetilde{G} e^{-2 \left| X\right| }+\mathcal{O}\left(\widetilde{G}^2\right)$. The local spectral density reads as
\begin{align}
n(x,\omega)=&2n_0\left(1-\sqrt{\gamma}\frac{G}{g} e^{-2 \left| x\right|/\xi }\right)\delta(\omega)
+\frac{1}{v}I\left(\frac{x}{\xi},\frac{\omega\hbar}{gn_0}\right).
\end{align}
Here $I(X,\Omega)$ is given by \eq{IG}. Note that even a weakly coupled impurity causes oscillations in the local spectral density for nonzero $\omega$.

At the position of the impurity $x=0$, the two contributions out of three that are proportional to the coupling strength $\widetilde{G}$ in \eq{IG} vanish, and we are left with the last term. The sign of this term is actually $-\text{sgn}({G})$, and thus for a repulsively interacting impurity ($G>0$) the local spectral density $n(0,\omega)$ decreases both for positive and for negative  $\omega$ while it increases for an attractive impurity ($G<0$). For nonzero position $x$, the impurity contribution in $n(x,\omega)$ oscillates and changes its sign as a function of $\omega$. 

For small frequency $|\Omega|\ll1$ we recover the expansion of \eq{IsmallOmega} in the limit $|\widetilde{G}|\ll1$. In the limit of high energies, the term proportional to $\widetilde{G}$ in \eq{IG} is of higher order in $1/\sqrt{|\Omega|}$ with respect to Eqs.~(\ref{largeOmegaN}) and (\ref{largeOmegaP}).

%%%%%%%%%%%%%%%%%%%%55
\section{Arbitrary coupling $\widetilde{G}$ and arbitrary frequency \label{sec:arbitrary}}
%%%%%%%%%%%%%%%%%%%%%%%

\begin{figure}
\includegraphics[width=\columnwidth]{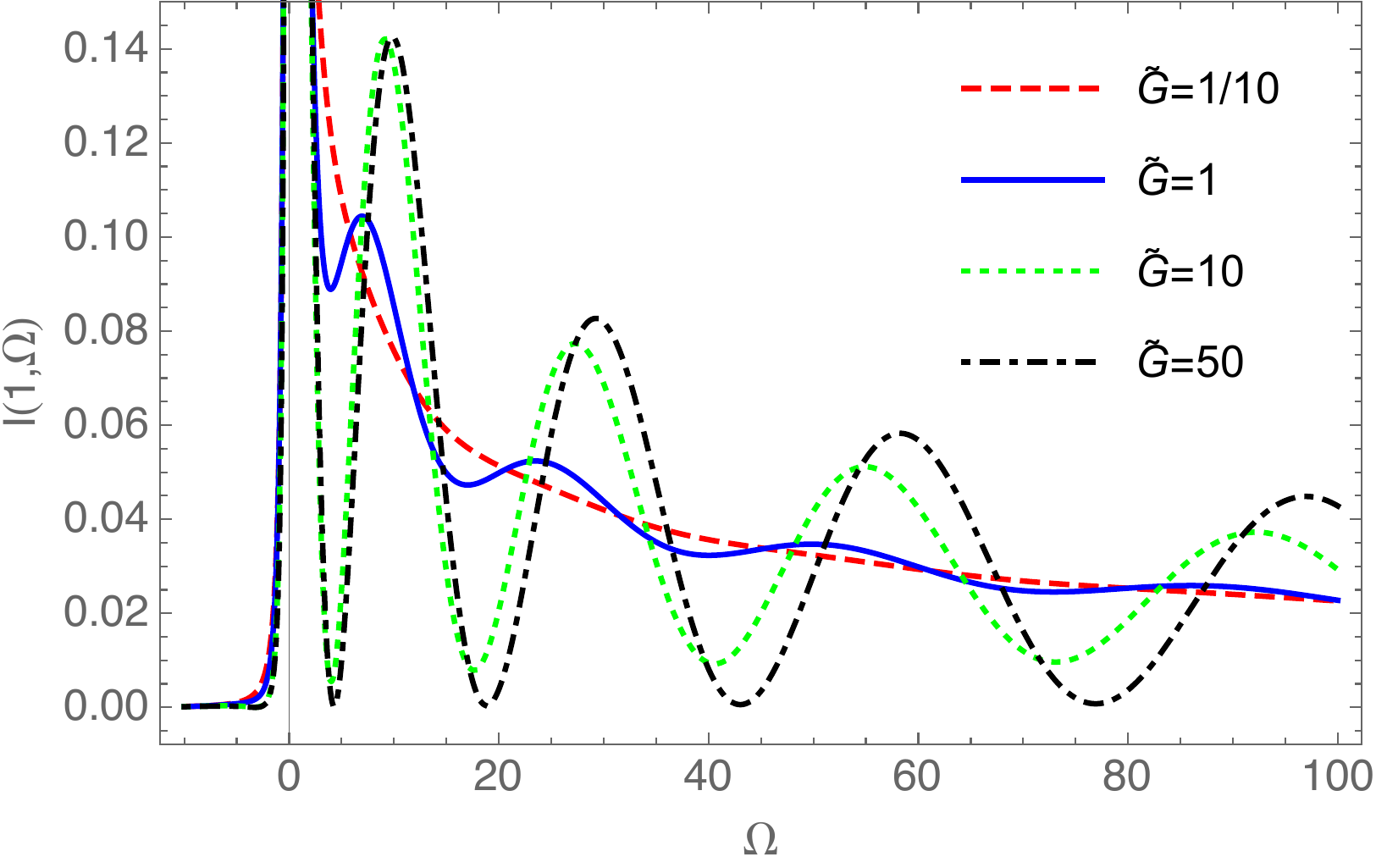}
\caption{(Color online) Dependence of $I(X,\Omega)$ on $\Omega$ for the fixed value $X=1$ and the temperature $T=\mu/10$ for different impurity coupling strengths.}
\label{Fig:IOmega}
\end{figure}

\begin{figure}
\includegraphics[width=\columnwidth]{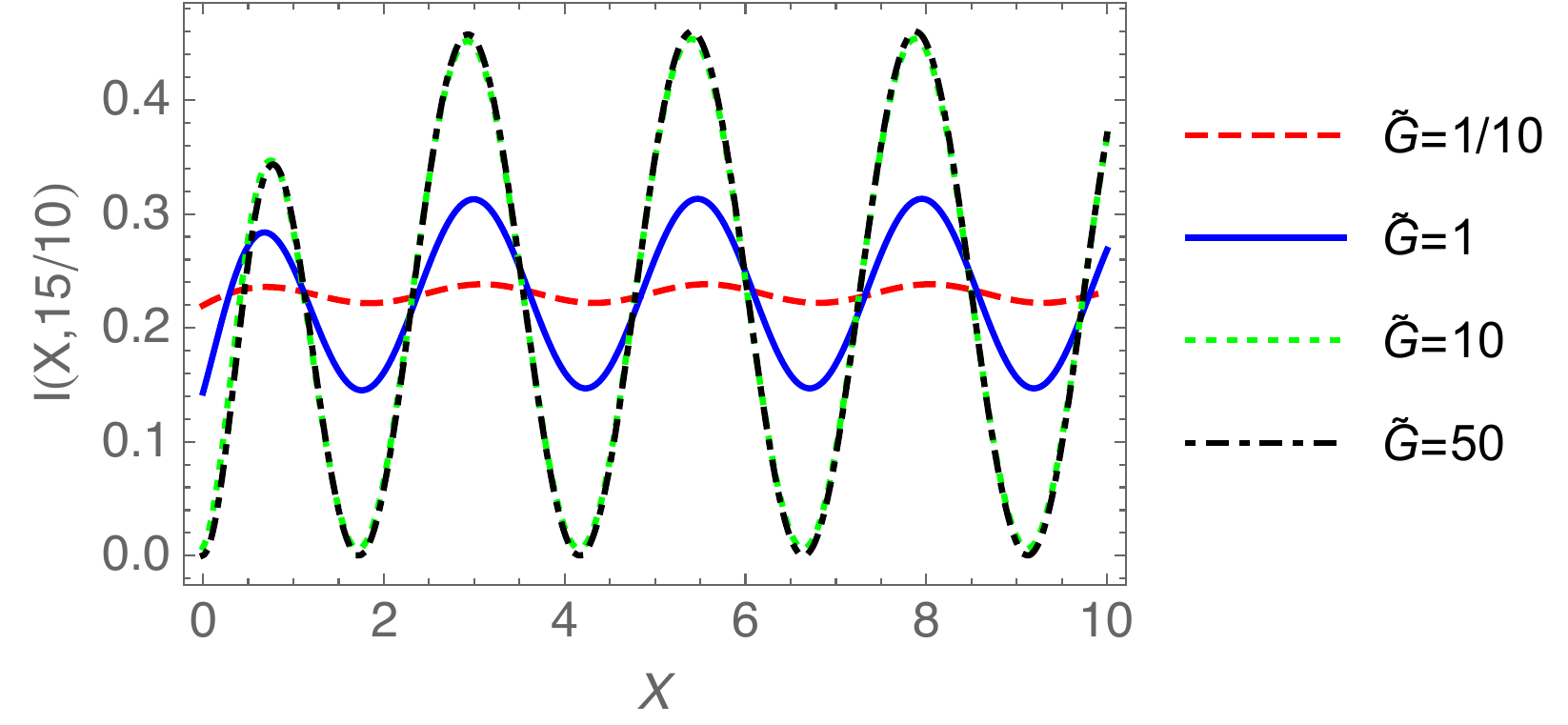}
\caption{(Color online) Dependence of $I(X,\Omega)$ on $X$ for the fixed value of $\Omega=15/10$ and the temperature $T=\mu/10$ for different impurity coupling strengths. }
\label{Fig:comparison}
\end{figure}

In this section we consider the local spectral density for an arbitrary impurity strength $\widetilde{G}$ and an arbitrary frequency. 
First we study the local spectral density far from the impurity $|x|\gg \xi$. There, the functions $S_i$ and $D_i$ with $i=3,4$ are approximately zero since they vanish exponentially rapidly (c.f.~\eq{S} and \eq{D}). This leads to $|u_k(X)|^2=|t u^{\mathrm{hom}}_k |^2$ and $|u_{-k}(X)|^2=|u^{\mathrm{hom}}_k|^2 |1+re^{2ikX}(2-ik)/(2+ik)|^2 $ for $k>0$ and 
$X>0$.  Here $u^{\mathrm{hom}}$ is the solution in the absence of the impurity, and thus it is $X$-independent. It is  given by \eq{homogeneU}. The same expressions hold for $v$ functions once $u$ and $u^{\mathrm{hom}}$ are replaced by $v$ and $v^{\mathrm{hom}}$, respectively. Using \eq{Iw} and $|r|^2+|t|^2=1$, we obtain 
\begin{align}\label{BigX}
I(X,\Omega)=I^{\mathrm{hom}}(\Omega)\left[ 1+\left(r_{\Omega} e^{2 i k_{\Omega} |X|}\frac{2-i k_{\Omega}}{2(2+i k_{\Omega})}+\mathrm{h.c.}\right)\right],
\end{align}  
for $|X|\gg 1$. Here $\mathrm{h.c.}$ stands for the hermitian conjugate term, $I^{\mathrm{hom}}(\Omega)$ is given by \eq{additional} and $r_{\Omega}$ is $r$ given by \eq{r} and evaluated for $k=k_{\Omega}$. Note that the dependence of $I(X,\Omega)$ on $\widetilde{G}$ is carried by $r_{\Omega}$, that takes values in the complex plane. 

The local spectral density is a periodic function of position far from the impurity. Its period does not depend on the coupling strength $\widetilde{G}$. It is determined by the inverse momentum of Bogoliubov quasiparticles $\pi\xi/k_{\Omega}$ that are probed at given energy $\Omega=\hbar\omega/\mu$. The average value of $ n(x,\omega)$ for any frequency and any impurity strength equals the local spectral density in the absence of the impurity, i.e., $\int_0^{\pi/k_{\Omega}}I(X,\Omega) dXk_{\Omega}/\pi=I^{\mathrm{hom}}(\Omega)$. Equation (\ref{BigX}) shows how the amplitude of oscillations far from the impurity depends on the dimensionless parameters: energy and the impurity strength.

Next we consider an arbitrary $X$. In Fig.~\ref{Fig:contour}, we present the position and the frequency dependance of $I(X,\Omega)$ given by \eq{Iw} for the impurity strength $\widetilde{G}=10$ and the temperature $T=\mu/10$. We notice the oscillations and the pronounced asymmetry for positive and negative frequency as discussed in previous sections. In Fig.~\ref{Fig:IOmega} we present $I(1,\Omega)$ for different impurity coupling strengths, showing that the amplitude of oscillations increases with $|\widetilde{G}|$. The same property is observed in Fig.~\ref{Fig:comparison} where $I(X,15/10)$ is shown for several values of $\widetilde{G}$. 
For repulsive impurity-bosons interaction, $I(X,\Omega)$ saturates to values given by \eq{Ib} for an infinitely strong coupling constant $\widetilde{G}$. We notice the decrease of $I(0,\Omega)$ as $\widetilde{G}$ is increased and its drop to zero value for strong coupling constant. 

%%%%%%%%%%%%%%%%%%%
\section{Summary}\label{sec:conclusions}
%%%%%%%%%%%%%%%%%%%
\begin{figure}
\includegraphics[width=\columnwidth]{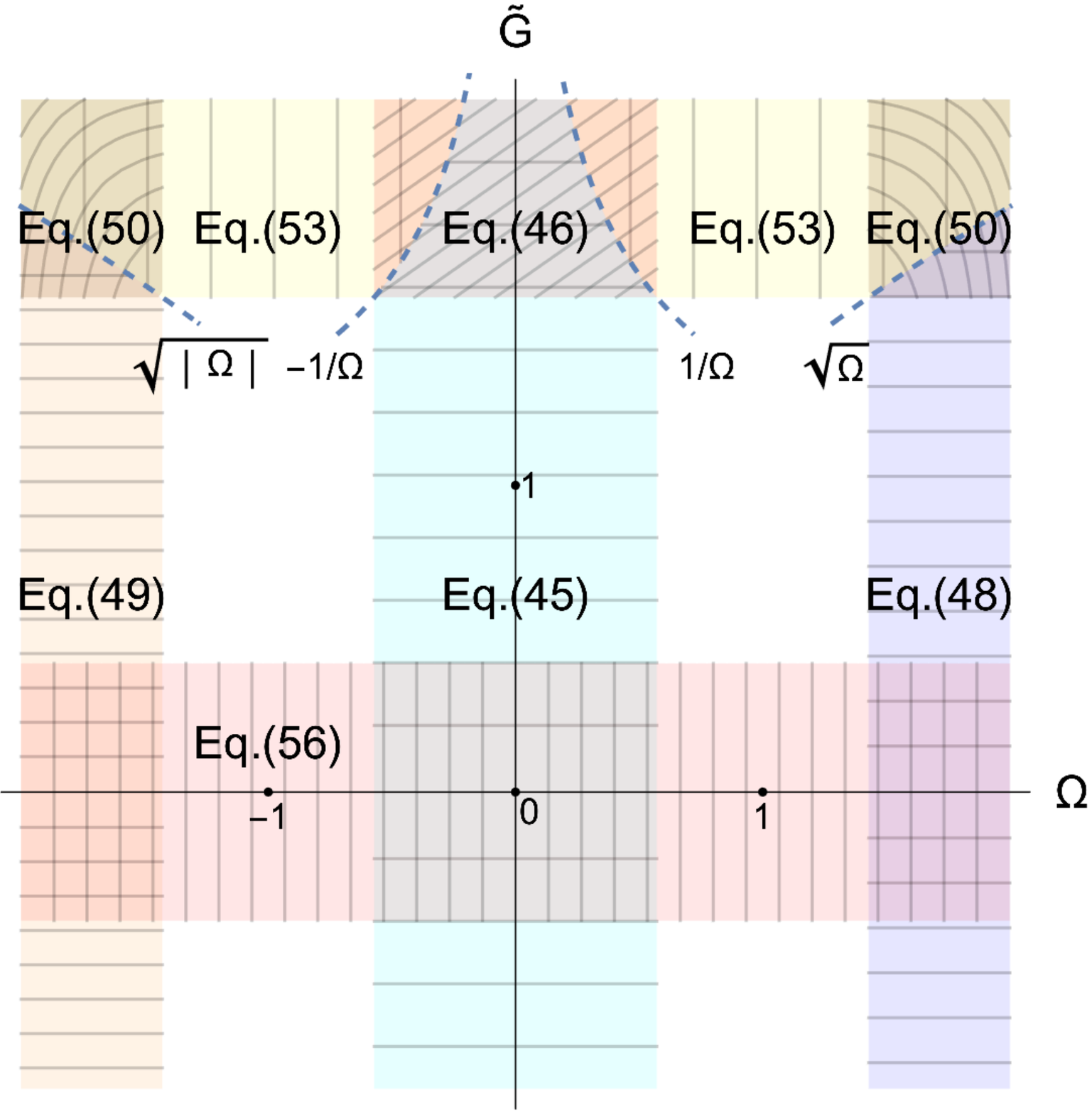}
\caption{(Color online) Schematic diagram representing the asymptotic behavior of $I(X,\Omega)$ given by \eq{Iw} in the $\widetilde{G}-\Omega$ plane. The local spectral density satisfies \eq{nwn}. Note that each equation has its own mesh that specifies the region of its applicability.}
\label{Fig:results}
\end{figure}

In this paper we have studied a system of 1D weakly interacting bosons in the presence of an impurity. 
%We have focused on the local spectral density, although the obtained results can be used for calculation of other correlation functions.  
We have analytically evaluated the local spectral density of bosons for any frequency, position and positive impurity strength. In some parameter regions the expression for $n(x,\omega)$ simplifies considerably, and we thus summarize different asymptotic behaviors in Fig.~\ref{Fig:results}. 

Contrary to the local density that does not show the Friedel oscillations \cite{PhysRevResearch.2.043104}, the local spectral density has an oscillatory behavior for any impurity coupling strength. This behavior is presented in Fig.~\ref{Fig:contour}. At separations from the impurity greater than the healing length, $n(x,\omega)$ simplifies and is given by \eq{BigX} for all energies.

At zero temperature in the limit of low energies, $\hbar|\omega|\ll \mu$, $n(x,\omega)$ behaves as $ 1/|\omega|$. This singular behavior in $\omega$ persists for any impurity strength $G$, including the case of an impurity with an infinitely strong coupling constant that acts as a boundary, leading to complete reflection of Bogoliubov quasiparticles. The frequency dependence gets multiplied by a prefactor that describes the modification of the local spectral density due to the impurity and introduces the spatial dependence. For not too strong impurity coupling constant, the prefactor behaves as $\tanh^2{(|x|/\xi+X_0)}$, see \eq{SmallOmega1}. Here $x$ denotes the separation from the impurity, $\xi$ is the healing length and $X_0$ depends in a nonlinear way on the impurity coupling constant, the interaction strength between the bosons, their density and mass as given by \eq{xo}.
In the limit of high energies, $\hbar|\omega|\gg \mu$, and for any $G$, the local spectral density  scales as $1/|\omega|^{5/2}$ and as $1/\sqrt{\omega}$ for negative and positive energies, respectively. For not too high impurity strength, this result is multiplied by $\tanh^4{(|x|/\xi+X_0)}$ at negative energies, while for positive energies the impurity influence is not visible in the leading order result, see Eqs.~(\ref{largeOmegaP}) and (\ref{largeOmegaN}). In the strong coupling regime ($\widetilde{G}\gg1$), the oscillations in $n(x,\omega)$ are observable also at low and high energies, see \eq{Ib}. The effects of thermal fluctuations  in the local spectral density at  low temperatures are taken into account through the multiplicative factor $\coth{(\hbar \omega/2T)}$. Note that by increasing the interaction strength between the bosons, the power-law exponents of the frequency-dependence of the local spectral density in the bulk and in the vicinity of an impurity with infinitely strong positive coupling constant become different at low energies \cite{Cazalilla2}.

%%%%%%%%%%%%%%%%%
\section{Acknowledgements}
%%%%%%%%%%%%%%%%%

We acknowledge Z. Ristivojevic and B. Reichert for useful discussions, and E. Tiesinga and an unknown reviewer for a critical reading of the manuscript.

\appendix

%%%%%%%%%%%%%%%%%%%%%%%%%%%%%%%%%%
\section{Low-frequency dependence of $n(x,\omega)$ \label{AppA}}
%%%%%%%%%%%%%%%%%%%%%%%%%%%%%%%%%%

In this Appendix, we provide the calculation of the local spectral density  of interacting bosons in the thermodynamic limit in a homogeneous system (i.e.,~$G=0$) using the quadratic Luttinger liquid theory \cite{Haldane}. The results are valid for an arbitrary interaction strength but are limited to low energies. Note that for bosons with contact interaction, the Luttinger liquid parameter satisfies $K\geq 1$. 

We start by expressing the single particle operator as $\hat{\Psi}^{\dag}(x,t)=\sqrt{\hat{n}(x,t)}e^{i \hat{\theta}(x,t)}$. Then, the time ordered correlation function can be written as 
\begin{align}\label{correlation}
&\left\langle  \mathcal{T} \left[\hat\Psi(x,t)\hat\Psi^\dag(x,0)\right]\right\rangle\approx n_0\left\langle\mathcal{T}e^{-i\hat{\theta}(x,t)}e^{i\hat{\theta}(x,0)} \right\rangle
\\&=n_0  e^{-F(t)/2}.\label{theta-theta}
\end{align}
Here we approximated $\hat{n}$ by the mean density $n_0$, since the phase fluctuations give the dominant contribution to the correlation function at long times.
At finite temperatures, the correlation function \eq{theta-theta} takes the form \cite{Giamarchi} 
\begin{align}\label{finiteT}
&F(t)=\frac{1}{2K} \ln{\left\{  - \frac{v^2\hbar^2}{\pi^2\Lambda^2 T^2} \sinh^2{\left[\frac{\pi T }{\hbar v}\left(v |t| -i \epsilon\right)\right]} \right\}}
\end{align}
for $v |t|\gg \Lambda$. 
Here we implicitly assume the limit $\epsilon\to 0^+$.
We introduced the smooth cutoff $e^{-\Lambda |k|/\hbar}$ in the momentum $k$ space. It can be estimated as $\Lambda^{-1} \approx\mu/\hbar v$, such that only the linear part of the excitation spectrum is taken into account. Here $v$ denotes the sound velocity. The time-ordered correlation function at zero temperature simplifies to 
\begin{align}\label{zeroT}
F(t)=\frac{1}{2K}\ln{\left[\frac{-(v| t|-i\epsilon)^2}{\Lambda^2}\right]}.
\end{align}

Now we can evaluate the local spectral density given by \eq{nwdef}. The correlation function is
\begin{align}
\left\langle  \mathcal{T} \left[\hat\Psi(x,t)\hat\Psi^\dag(x,0)\right]\right\rangle=n_0\left(\frac{\Lambda}{v |t|}\right)^{\frac{1}{2K}}e^{-i \frac{\pi}{4K}}
\end{align}
in the leading order at zero temperature. 
After performing the Fourier transform with respect to time  we get
\begin{align}\label{eq:app}
n(\omega)=n_0\frac{\Gamma(1-1/2K)}{\pi}\left(\frac{|\omega|\Lambda}{v}\right)^{\frac{1}{2K}}
\frac{1}{|\omega|}\sin{[\pi/(2K)]},
\end{align}
where $\Gamma(x)$ denotes the Gamma function. We assumed $K>1/2$, as well as that $K$ is finite. In the limit of weakly interacting bosons, $K=\pi/\sqrt{\gamma}\gg 1$, the expression (\ref{eq:app}) becomes
\begin{align}\label{zeroTLL}
n(\omega)=n_0\frac{\sqrt{\gamma}}{2\pi| \omega|}.
\end{align} 
This result coincides with the low energy limit $\hbar|\omega|\ll g n_0$ of $\sqrt{\gamma}\frac{\hbar}{g}I^{\mathrm{hom}}(\omega\hbar/g n_0)$ where $I^{\mathrm{hom}}$ is given by \eq{additional}. Next we consider influence of thermal fluctuations. The correlation function reads as
\begin{align}
\left\langle  \mathcal{T}\left[\hat\Psi(x,t)\Psi^\dag(x,0)\right]\right\rangle=n_0\left(\frac{\pi\Lambda T}{\hbar v \sinh{(\pi T |t|/\hbar)}}\right)^{\frac{1}{2K}}e^{-i\frac{\pi}{4K}}.
\end{align}
We evaluate the local spectral density to be
\begin{align}\label{nwanyg}
n(\omega)=&\frac{\hbar n_0}{\pi^2 T}\left(\frac{2\pi \Lambda T}{\hbar v}\right)^{1/(2K)}\Gamma{\left( 1-\frac{1}{2K}\right)}\cos{\left( \frac{\pi}{4K}\right)}\notag\\ &\times\text{Re}\left[\frac{\Gamma{\left(\frac{1}{4K}-i\frac{\hbar \omega}{2\pi T}\right)}}{\Gamma\left(1-i\frac{\hbar\omega}{2\pi T}-\frac{1}{4K}\right)}\right].
\end{align}
In the limit of weakly interacting bosons, $K\gg 1$, this expression simplifies into
\begin{align}
n(\omega)=\sqrt{\gamma}\frac{n_0}{2\pi\omega}\coth{\left(\frac{\hbar \omega}{2T}\right)}.
\end{align}
This is the limit of  $\sqrt{\gamma}\frac{\hbar}{g}I^{\textrm{hom}}\left({\omega\hbar}/{gn_0}\right)$  for $\hbar|\omega|\ll  g n_0$ (cf.~\eq{additional}). 
For $\hbar|\omega|\ll T$, \eq{nwanyg} simplifies to  $n(\omega)=\sqrt{\gamma}\frac{n_0 T}{\pi \hbar\omega^2}$. 
In the opposite case, for $\hbar|\omega|\gg T$, we recover  the zero temperature result (\ref{zeroTLL}).
We note that the particle-hole symmetry that characterizes the Luttinger liquid theory is reflected in the fact that the local spectral density is an even function of $\omega$.  

Let us return to the correlation function (\ref{correlation}) and compare the approach of this section with the one in the main text. We are now interested in weakly interacting bosons only, i.e., $\gamma\ll 1$. 
In the case of the finite system size $L$, the correlation function $F(t)$ in \eq{theta-theta} saturates into a finite small constant  for long times $|t|\gg L/v$ provided the system size is smaller then the phase coherence length $L\ll \ell_{\phi}$. Thus the phase fluctuations are always small $F(t)\ll 1$. This implies the existence of the condensate.  We can thus expand the exponential in \eq{theta-theta} and perform the Fourier transform. We obtain the $2 n_0 \delta(\omega)$ term in \eq{nwn} from the leading order term. The next order term $\sim\sqrt{\gamma}$ is determined by the Fourier transform of $F(t)$. To evaluate it at low nonzero frequency we can use (\ref{finiteT}), which is valid for $\hbar/ \mu \ll |t|\ll L/v$, leading to the low frequency limit of the finite frequency dependence in \eq{nwn} in the homogeneous case. At higher frequencies one has to take into account nonlinearities of the quasiparticle spectrum as well as the fluctuations of the density. 

More generally, i.e., beyond the low-energy behavior that we discussed above, as well as beyond the homogeneous case, we can compare the two description, the one in the main text based on the expansion \eq{expansion} and the one based on the phase-density representation of the field operator. We point out that the phase and the density can be related to the field operators as \cite{PhysRevA.67.053615,Casimir}
\begin{align}\label{relation}
\hat\psi_1(X,\tau)=\frac{\delta \hat{n}(X,\tau)}{2\psi_0(X)}-i\psi_0(X)\hat\theta_1(X,\tau)
\end{align}
where the hermitian operator $\gamma^{1/4}\mu\delta \hat{n}(X,\tau)/g$ denotes the leading order correction to the mean-field density $\mu\psi_0^2(X)/g$. The phase is $\theta(X\xi_{\mu},\tau\hbar/\mu)=\gamma^{1/4}\hat\theta_1(X,\tau)+\tau$ in the leading order.  We use the dimensionless position $X$ and time $\tau$ defined in Sec.~\ref{model}. The expression (\ref{relation}) and its conjugate allow us to express  the phase and the density as a function of the field operators, and then using the results for $\psi_0$ and $\psi_1$ in the main text obtain the expressions for $\hat{n}$ and $\hat{\theta}$ at any impurity coupling strength $G$. 

At low energies one can use the Luttinger-liquid description for a system of bosons also in the presence of the boundaries \cite{Cazalilla2}. It was shown in Ref.~\cite{Cazalilla2} that the exponent of the frequency in the local single-particle density of states in the vicinity of the boundary (or an infinitely repulsive impurity) differs from the bulk one that we have discussed above. The exponent is given by $-1+1/K$ at zero temperature \cite{Cazalilla2}. 

%%%%%%%%%%%%%%%%%%%%%%%%%%%%%%%%%%%
\section{Homogeneous system: $\tilde{G}=0$}\label{App:homogene}
%%%%%%%%%%%%%%%%%%%%%%%%%%%%%%%%%%%

In the absence of the impurity, the results from Sec.~\ref{sec:solution} simplify and one obtains $\psi_0^{\textrm{hom}}=1$, 
\begin{align}\label{homogeneU}
u_k^{\textrm{hom}}(X)=\frac{2 + i |k|}{2}e^{i k X}\left(1+\frac{k^2}{2\epsilon_k}\right),\\\label{homogeneV}
v_k^{\textrm{hom}}(X)=\frac{2 + i |k|}{2}e^{i k X}\left(1-\frac{k^2}{2\epsilon_k}\right).
\end{align}

%%%%%%%%%%%%%%%%%%%%%
\section{Low energy limit}\label{LowFrequency}
%%%%%%%%%%%%%%%%%%%%%%%%

In this section we provide expressions for the functions $u_k$ and $v_k$ at small $|k|\ll 1$, which follow from the results obtained in Sec.~\ref{sec:solution}. The coefficients become $t = -1$, $r = 0$ and $r_e=i k e^{-2X_0}/2$ to leading order. The functions read as
\begin{align}\label{usmallw}
u_k(X)=& \frac{1}{2} \text{sech}^2(| X| +X_0) \Big\{e^{i k X} [\sinh (2 | X|+2X_0
   )\notag\\ &-i \text{sgn}(X)]+i \text{sgn}(X)\Big\}
\end{align}
and 
\begin{align}\label{vsmallw}
v_k(X)=& \frac{1}{2} \text{sech}^2(| X| +X_0) \Big\{e^{i k X} [\sinh (2 | X|+2X_0
   )\notag\\ &+i \text{sgn}(X)]-i \text{sgn}(X)\Big\}
\end{align}
for small positive $k\ll 1$ and arbitrary $X$. For negative $k$, we easily obtain the expressions by using the relation
\begin{align}\label{characteristic}
u_k(X)=u_{-k}(-X)
\end{align}
explained in Sec.~\ref{sec:solution}. The same property holds for $v_k(X)$. Note that the functions (\ref{usmallw}) and (\ref{vsmallw}) are continuous function of $X$, as has to be the case (see Sec.~\ref{sec:solution}).

%%%%%%%%%%%%%%%%%%%%%%%%%%%%%%%%%%%%%%
\section{Large coupling $\widetilde{G}\gg 1$ and low energy $\hbar|\omega|\ll\mu$}\label{app:BigGSmallOmega}
%%%%%%%%%%%%%%%%%%%%%%%%%%%%%%%%%%%%%%

In this subsection we consider the limit of low energy and large positive $\widetilde{G}$ such that  $\widetilde{G}^{-1}\sim|k| \ll 1$.
We evaluate to leading order
\begin{align}
u_k(X)=\frac{\text{sech}{(X)}^2 }{2 (i+\widetilde{G} k)}\left\{-1+e^{i k X} [1+i \sinh (2 X)]\right\}
\end{align}
for small positive $k\ll 1$ and $X>0$, while for 
$X<0$ we get
\begin{align}
u_k(X)
%=&-e^{i k X} \tanh (X)+\frac{\text{sech}^2(X) }{2 (i+\widetilde{G} k)}\Big\{ (i G k-1) e^{i k X}\notag\\&-\widetilde{G} k e^{-i k X}
%[\sinh (2 X)+i]+1\Big\}.\\
   =&-e^{i k X} \tanh (X)+i\frac{\text{sech}^2(X) }{2}  e^{i k X}\notag\\&+\frac{\text{sech}^2(X) }{2 (i+\widetilde{G} k)}\Big\{-\widetilde{G} k e^{-i k X}
   [\sinh (2 X)+i]+1\Big\}.
\end{align}
For small positive $k\ll 1$ and $X>0$, the function $v_k(X)$ reads as
\begin{align}
v_k(X)=\frac{\text{sech}{(X)}^2 }{2 (i+\widetilde{G} k)}\left\{1+e^{i k X} [-1+i \sinh (2 X)]\right\},
\end{align}
while for $X<0$ it has the form
\begin{align}
v_k(X)=&-e^{i k X} \tanh (X)+\frac{\text{sech}^2(X)}{2 (i+\widetilde{G} k)}\Big\{-i(i+\widetilde{G}k)e^{i k X}\notag\\&+\widetilde{G} k e^{-i k X}
   [-\sinh (2 X)+i]-1\Big\}.
\end{align}

%%%%%%%%%%%%%%%%%%%%%%%%%
\section{High energy limit $\hbar|\omega|\gg \mu$}\label{HighFrequency}
%%%%%%%%%%%%%%%%%%%%%%%%%%

We obtain 
\begin{align}
u_k(X)=i k e^{i k X},
\end{align}
and
\begin{align}
v_k(X)=\frac{i e^{i k X} }{k}\tanh ^2(|X|+X_0),
\end{align}
for $k\gg 1$.

%%%%%%%%%%%%%%%%%%
\section{Large coupling $\tilde{G}\gg 1$ and high energy limit $\hbar|\omega|\gg \mu$}\label{HighFrequencyB}
%%%%%%%%%%%%%%%%%%%

We evaluate $u_k$ for large positive $k\sim \widetilde{G}$ and $X<0$ to be 
\begin{align}
u_k(X)= \frac{\tilde G k}{k+i \tilde G}e^{-i k x}+i k e^{i k x},
\end{align}
and for $X>0$ it reads as
\begin{align}
u_k(X)=\frac{k^2 e^{i k x}}{\widetilde{G}-i k}
\end{align}
to leading order. 
We evaluate also 
\begin{align}
v_k(X)=\frac{e^{i k X} }{\tilde G-i k}\tanh ^2(X)
\end{align}
for $X>0$, while for $X<0$ it takes the form 
\begin{align}
v_k(X)=\frac{ -2 \tilde G \sin (k X)+k e^{i k
   X}}{k (\tilde G-i k)}\tanh ^2(X)
\end{align}
to leading order.

%%%%%%%%%%%%%%%%%%%%%%%%%%%%
\section{Infinitely strong impurity coupling constant $\widetilde{G}=\infty$}\label{AppInfinitely}
%%%%%%%%%%%%%%%%%%%%%%%%%%%%

Here, we consider the case of $\widetilde{G}=\infty$. The condensate wave-function takes the form
\begin{align}
\psi_0(X)=&\tanh {|X|} \label{p0},
\end{align}
and  for $X\geq0$ and $k<0$ \cite{PhysRevResearch.2.043104} we find 
\begin{align}
u_k(X)= & k\left[1+ \dfrac{k^2+2\cosh^{-2}X}{2\epsilon_k} \right]\sin(kX) \notag\\
	&+2\left[1+ \dfrac{k^2}{2\epsilon_k}\right]\cos(k X)\tanh X,\label{uk}\\
v_k(X)= &k\left[1- \dfrac{k^2+2\cosh^{-2}X}{2\epsilon_k} \right]\sin(kX) \notag\\
	&+2\left[1- \dfrac{k^2}{2\epsilon_k}\right]\cos(k X)\tanh X.\label{vk}
\end{align}
For $X\geq0$ and $k>0$ the functions $u_k$ and $v_k$ vanish. The impurity imposes vanishing transmission coefficient, thus leading to $t=t_e=r_e=0$ and $r=1$ in \eq{solution}, see Eqs.~(\ref{r})--(\ref{Re}). One easily calculates the functions $u_k(X)$  and $v_k(X)$ for negative $X$ using the property (\ref{characteristic}) and the above solutions (\ref{uk}) and (\ref{vk}).

%%%%%%%%%%%%%%%%%%%%%%%%%%%%
\section{Weak coupling $|\widetilde{G}|\ll 1$\label{app:weakly}}
%%%%%%%%%%%%%%%%%%%%%%%%%%%%

In this section we consider the effects of a weakly coupled impurity, $|\widetilde{G}|\ll 1$. Here we provide expressions for $u_k$ and $v_k$ obtained using the results from Sec.~\ref{sec:solution}. In the limit of interest, one should replace $\eta= 1-\widetilde{G}/2 +\mathcal{O}(\tilde G^2)$ in the coefficients given by Eqs.~(\ref{r}),  (\ref{t}), and (\ref{Re}). For dimensionless coordinate $X\geq 0$ and momentum $k>0$ we obtain
\begin{align}
 \label{eq:1}
u_k(X)=&u_k^{\textrm{hom}}(X)+\frac{\widetilde{G}}{2(k+2
   i) \left(k^2+2\right)} \Big[
   - e^{i k X} k q (k+q)\notag\\&+e^{-2 X+i k X}\left(k^2+2\right) (k+q-2 i)\notag\\&+2e^{-q X}(k-q)\Big]
\end{align}
in the first order in $\widetilde{G}$.
%\begin{align}
% \label{eq:1}
%u_k(X)=u_k^{\textrm{hom}}(X)+\widetilde{G} \frac{
   %- e^{i k X} k q (k+q)+e^{-2 X+i k X}\left(k^2+2\right) (k+q-2 i)+2e^{-q X}(k-q)}{2(k+2
   %i) \left(k^2+2\right)}+O\left(\widetilde{G}^2\right).
%\end{align}
Here $q=\sqrt{4+k^2}$, while  $u_k^{\textrm{hom}}(X)$ denotes solution (\ref{homogeneU}) obtained in the homogeneous case (i.e.~the case $\widetilde{G}=0$). For $X\leq 0$ and $k>0$
\begin{align}\label{eq:2}
u_k(X)=&u_k^{\textrm{hom}}(X)-\frac{\widetilde{G}}{2 (k+2 i)
   \left(k^2+2\right)}\Big[
   -e^{-i k X}k q (k+q)\notag\\& +e^{2 X+i k X}\left(k^2+2\right) (k+q+2 i) +2e^{q
   X} (k-q)\Big].
\end{align}
%\begin{align}\label{eq:2}
%u_k(X)=u_k^{\textrm{hom}}(X)-\widetilde{G}\frac{
  % -e^{-i k X}k q (k+q) +e^{2 X+i k X}\left(k^2+2\right) (k+q+2 i) +2e^{q
   %X} (k-q)}{2 (k+2 i)
   %\left(k^2+2\right)}+O\left(\widetilde{G}^2\right).
%\end{align}
For negative $k$ values we easily obtain $u_k$ and $v_k$ by using the relation (\ref{characteristic})
and Eqs.~(\ref{eq:1}) and (\ref{eq:2}). The same property holds for $v_k(X)$. As expected, in the presence of an impurity we have $|u_k(X)|\neq|u_{-k}(X)|$ due to scattering of the quasiparticles at the impurity.
Next we evaluate the functions describing hole--like excitations. For $X\geq 0$ and $k>0$ they read as 
\begin{align}\label{eq:3}
v_k(X)=&v_k^{\mathrm{hom}}(X)+\frac{\widetilde{G}}{2
   (k+2 i) \left(k^2+2\right)}\Big[-e^{i k X}k q(q-k) \notag\\&+e^{-2 X+i k X}\left(k^2+2\right) (k-q-2 i) \notag\\&+2 e^{-q X} (k+q)\Big].
\end{align}
%\begin{align}
%v_k(X)=v_k^{\mathrm{hom}}(X)+\widetilde{G}\frac{-e^{i k X}k q(q-k) +e^{-2 X+i k X}\left(k^2+2\right) (k-q-2 i) +2 e^{-q X} (k+q)}{2
  % (k+2 i) \left(k^2+2\right)}+O\left(\widetilde{G}^2\right)
%\end{align}
Here $v_k^{\mathrm{hom}}(X)$  stands for the solution  in the absence of the impurity (\ref{eq:2}).
For $X\leq 0$ and $k>0$ we obtain
\begin{align}\label{eq:4}
v_k(X)=&v_k^{\mathrm{hom}}(X)-\frac{\widetilde{G}}{{2 (k+2 i)
   \left(k^2+2\right)}}\Big[
  -e^{-i k X}k q(q-k) \notag\\&+e^{2 X+i k X}\left(k^2+2\right) (k-q+2 i)
   +2 e^{q X}(k+q) \Big].
\end{align}
%\begin{align}
%v_k(X)=i e^{i k X}\frac{ k (k-q)+4 }{2 (k+2 i)}-\widetilde{G}\frac{
 % -e^{-i k X}k q(q-k) +e^{2 X+i k X}\left(k^2+2\right) (k-q+2 i)
   %+2 e^{q X}(k+q) }{2 (k+2 i)
   %\left(k^2+2\right)}+O\left(\widetilde{G}^2\right).
%\end{align}
Notice that the expressions (\ref{eq:3}) and (\ref{eq:4}) for $v_k(X)$ can be obtained from the expressions (\ref{eq:1}) and (\ref{eq:2}) for $u_k(X)$ after replacing $q$ by $-q$ everywhere, except in the exponential $e^{ -q |X|}$. This explains the difference between the positive and the negative frequency dependence in \eq{IG} given by the exponential $\exp[-2|\Omega X|/k_{\Omega} ]$.

%\bibliography{bib}
%\bibliographystyle{apsrev4-1}
\end{document}